 \documentclass[prb,twocolumn,showpacs,showkeys]{revtex4}

\usepackage{graphicx}
\usepackage{amssymb,amsmath}
\usepackage{dcolumn}
\usepackage{bm}
\usepackage{color}

\begin{document}

\title{The Fermi surface of $ {\bf MoO_2} $ as studied by angle-resolved
       photoemission spectroscopy, de Haas-van Alphen measurements, and
       electronic structure calculations}

\author{Judith Moosburger-Will}
\affiliation{Experimentalphysik II, Institut f\"ur Physik,
             Universit\"at Augsburg, D-86135 Augsburg, Germany}
\author{J\"org K\"undel}
\affiliation{Experimentalphysik II, Institut f\"ur Physik,
             Universit\"at Augsburg, D-86135 Augsburg, Germany}
\author{Matthias Klemm}
\affiliation{Experimentalphysik II, Institut f\"ur Physik,
             Universit\"at Augsburg, D-86135 Augsburg, Germany}
\author{Siegfried Horn}
\affiliation{Experimentalphysik II, Institut f\"ur Physik,
             Universit\"at Augsburg, D-86135 Augsburg, Germany}
\author{Philip Hofmann}
\affiliation{Institute for Storage Ring Facilities (ISA) and
             Interdisciplinary Nanoscience Center (iNANO),
             University of Aarhus, DK-8000 Aarhus C, Denmark}
\author{Udo Schwingenschl\"ogl}
\affiliation{ICCMP, Universidade de Bras\'ilia, 70904-970 Bras\'ilia, DF,
             Brazil and \\
             Theoretische Physik II, Institut f\"ur Physik,
             Universit\"at Augsburg, D-86135 Augsburg, Germany}
\author{Volker Eyert}
\email[Corresponding author. \\ Email-address:]{eyert@physik.uni-augsburg.de}
\affiliation{Center for Electronic Correlations and Magnetism,
             Institut f\"ur Physik, Universit\"at Augsburg,
             D-86135 Augsburg, Germany}
\date{\today}

\begin{abstract}
A comprehensive study of the electronic properties of monoclinic
$ {\rm MoO_2} $ from both an experimental and a theoretical point
of view is presented. We focus on the investigation of the Fermi
body and the band structure using angle resolved photoemission
spectroscopy, de Haas-van Alphen measurements, and electronic
structure calculations. For the latter, the new full-potential
augmented spherical wave (ASW) method has been applied. Very good
agreement between the experimental and theoretical results is found.
In particular, all Fermi surface sheets are correctly identified
by all three approaches. Previous controversies concerning additional
hole-like surfaces centered around the Z- and B-point could be resolved;
these surfaces were an artefact of the atomic-sphere approximation
used in the old calculations. Our results underline the importance
of electronic structure calculations for the understanding of
$ {\rm MoO_2} $ and the neighbouring rutile-type early transition-metal
dioxides.  This includes the low-temperature insulating phases of
$ {\rm VO_2} $ and $ {\rm NbO_2} $, which have crystal structures
very similar to that of molybdenum dioxide and display the well-known
prominent metal-insulator transitions.
\end{abstract}



\maketitle

\section{Introduction}
\label{intro}

The metal-insulator transitions of the early transition-metal oxides
have been attracting a lot of attention for decades. In particular,
since the first report by Morin, \cite{morin59} much interest has
centered about the cornerstone materials $ {\rm V_2O_3} $ and
$ {\rm VO_2} $. This is mainly due to the fact, that these compounds
display first order phase transitions with very narrow hystereses
of only few Kelvin and strong changes of the conductivity of several
orders of magnitude, making them promising candidates for applications.
Interestingly, the transitions of both compounds are accompanied by
characteristic structural changes, which initiated a discussion about
the driving forces. Dispute centered about the question whether the
transitions are driven predominantly by structural instabilities
of the parent corundum and rutile structure, respectively, or else
by strong electronic correlations. Nowadays, the importance of the
latter is widely accepted. However, despite intense work, the
questions concerning the origins of the respective transitions have
not yet been satisfactorily answered.

In $ {\rm VO_2} $, the metal-insulator transition occurs at 340\,K 
and is connected with a structural change from the rutile structure 
to a distorted monoclinic structure. The latter is characterized 
by (i) pairing of the metal atoms within chains parallel to the 
rutile $ {\bf c} $-axis and (ii) their lateral antiferroelectric 
zigzag-like displacement. \cite{andersson56,longo70} Electronic 
states near the Fermi energy are mainly of V $ 3d $ $ t_{2g} $
character. They fall into the $ d_{\parallel} $ band, which 
mediates V-V overlap along the metal chains, and the remaining, 
$ e_g^{\pi} $ bands. \cite{goodenough71b} At the transition, 
splitting of the $ d_{\parallel} $ band and upshift of the 
$ e_g^{\pi} $ bands due to increased metal-oxygen overlap
produce a finite band gap. Controversial discussions addressed the 
origin of the $ d_{\parallel} $ splitting, which was assigned either 
to the metal-metal dimerization or to increased electronic correlations 
resulting from the reduced screening by the $ e_g^{\pi} $ electrons.
\cite{goodenough60,adler67a,adler67b,goodenough71b,zylbersztejn75,rice94} 
State of the art electronic structure calculations gave strong hints at 
a structural instability but were not able to reproduce the insulating 
gap due to the shortcomings of the local density approximation (LDA).
\cite{wentzcovitch94,habil,eyert02b}  Only recently, LDA+DMFT 
calculations have demonstrated that the metal-insulator transition 
may be regarded as a correlation-assisted Peierls-like transition. 
\cite{biermann05}

\begin{figure}[htb]
\centering
\includegraphics[width=0.9\columnwidth,clip]{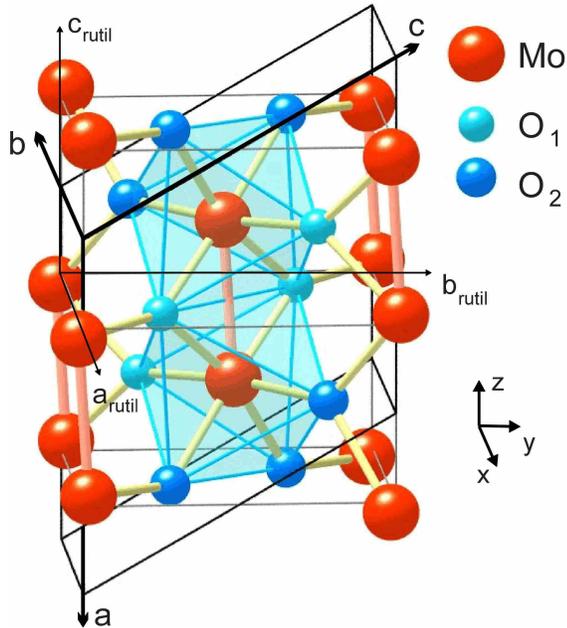}
\caption{(Color online) Monoclinic crystal structure of $ {\rm MoO_2} $.
         The dimerized metal pairs are highlighted by thick bonds.}
\label{introfig1}
\end{figure}

Remarkably, the structural distortions occurring at the transition 
of $ {\rm VO_2} $ are characteristic of most of the neighbouring 
$ d^1 $, $ d^2 $, and $ d^3 $ transition-metal dioxides. In particular, 
the metallic oxides $ {\rm MoO_2} $, $ {\rm WO_2} $, $ {\rm TcO_2} $, 
and $ \alpha $-$ {\rm ReO_2} $ all display the above-mentioned 
metal-metal pairing as well as the lateral displacement and crystallize 
in the same monoclinic structure as low-temperature $ {\rm VO_2} $. 
\cite{rogers69} This crystal structure is displayed in Fig.\ \ref{introfig1},
which allows to identify both the pairing and the lateral displacement
of the metal atoms.

In contrast to the above-mentioned metallic $ d^2 $ and $ d^3 $ members, 
the $ d^1 $ compound $ {\rm NbO_2} $, like $ {\rm VO_2} $, undergoes 
a metal-insulator transition (at 1081 K) and a simultaneous structural 
transition from rutile to a distorted variant having a body-centered 
tetragonal (bct) lattice. \cite{janninck66,sakata67} However, despite 
the differences in long-range order, the local deviations from the 
rutile structure are the same as in $ {\rm VO_2} $, i.e.\ the niobium 
atoms dimerize and experience lateral displacements at the transition. 
\cite{marinder62,cheetham76,pynn76}

Despite of the strong interest in the metal-insulator transition
of $ {\rm VO_2} $, attempts at understanding the origin of the common
structural characteristics are very rare. Indeed, any theory aiming
at the metal-insulator transition of $ {\rm VO_2} $ would be incomplete
without proper account of the local structural distortions of all the
rutile-related transition-metal dioxides. In order to close the
apparent gap, we have started a systematic investigation especially on
$ {\rm VO_2} $, $ {\rm NbO_2} $, and $ {\rm MoO_2} $ already a decade
ago, which by now provides a unified picture.
\cite{habil,eyert00,eyert02a,eyert02b}

Interest in $ {\rm MoO_2} $ was especially motivated by the metallic 
nature of this material, the reduced localization of the $ 4d $ 
electrons as compared to the $ 3d $ electrons of $ {\rm VO_2} $, 
and the rather small effective mass, which is similar to the free 
electron mass.  \cite{volskii71,volskii73,klimm97} These facts point 
at a rather weak influence of electronic correlations. 
Indeed, electronic structure calculations for $ {\rm MoO_2} $ show 
a large bonding-antibonding splitting of the $ d_{\|} $ bands. 
\cite{habil,eyert00} Since the occupied $ d_{\|} $ bands lie far 
below the Fermi level, a possible influence of electronic 
correlations in these bands would not lead to an enhanced 
effective mass or susceptibility. As a consequence, $ {\rm MoO_2} $
can be regarded as a ``model system'' for studying the origin of the 
structural characteristics of the early transition-metal dioxides. 
In addition, an extensive examination of the electronic structure 
of $ {\rm MoO_2} $ could help clarifying open questions of the 
above-mentioned structurally related systems, in particular of the 
$ d^1 $ members $ {\rm NbO_2} $ and $ {\rm VO_2} $.

In a previous work, we have presented a comprehensive theoretical 
study of the electronic properties of $ {\rm MoO_2} $. 
\cite{habil,eyert00} Electronic structure calculations using the 
ASW method, which is based on density functional theory and the 
local density 
approximation (LDA), were used to analyze the electronic states 
involved in low-energy excitations. In addition, emphasis was put 
on investigating the origin of the small but distinct deviations 
of the monoclinic crystal structure from the parent rutile structure, 
which for this material mainly consist of a metal-metal dimerization
parallel to the rutile ${ \bf c }$-axis. In short, we were able to 
show that the monoclinic structure results from what we called an 
embedded Peierls-like instability. This instability affects the 
quasi one-dimensional Mo $ 4d_{\parallel} $ bands, which mediate 
the metal-metal overlap along chains parallel to the rutile 
$ {\bf c } $-axis and which are embedded in a background of the 
remaining three-dimensionally dispersing Mo $ 4d $ $ e_g^{\pi} $,
electrons, which are also designated as the $ \pi^{\ast} $ electrons. 
The validity of this scenario had been demonstrated by 
comparing the electronic properties calculated for the monoclinic 
structure with those of a properly symmetrized artificial rutile 
structure. In addition, we found nearly perfect agreement of our 
calculated results with the outcome of ultraviolet photoemission 
spectroscopy and x-ray absorption spectroscopy apart 
from an underestimation of the $ d_{\parallel} $ band splitting by 
about 1\,eV due to the well known shortcomings of the LDA. 
\cite{habil,eyert00}

This scenario has also been shown to explain the metal-insulator 
transitions of the $ d^1 $ members $ {\rm VO_2} $ and 
$ {\rm NbO_2} $, which likewise are subject to Peierls-like 
instabilities of the quasi one-dimensional V $ 3d_{\parallel} $ bands 
in the embedding backgrounds of the $ \pi^{\ast} $ electrons. 
\cite{eyert02a,eyert02b} As in $ {\rm MoO_2} $, these instabilities
cause distortions of the rutile structure by the characteristic 
metal-metal dimerization. Due to the particular position of the Fermi 
energy in the $ d^1 $ compounds, the Peierls-like instabilities are then 
accompanied by the observed metal-insulator transitions. At the same 
time, the opening of the optical band gaps is supported by the lateral 
antiferroelectric displacements of the metal atoms, which cause an 
upshift of the $ \pi^{\ast} $ states.

Continuing our study of $ {\rm MoO_2} $, we concentrate in the present 
work especially on the Fermi surface and the band dispersions. 
For the first time, a comprehensive investigation of the full 
Fermi body of $ {\rm MoO_2} $, as determined both experimentally 
and theoretically, will be presented. In particular, we turn 
to a comparison of the theoretical results with angle-resolved 
photoemission spectroscopy (ARPES) and de Haas-van Alphen data 
(dHvA). In doing so, we apply a new implementation of the 
augmented spherical wave method, which goes beyond the 
version used in our previous work by taking into account the full 
potential rather than the approximate muffin-tin potential. The 
resulting changes of the electronic states are small but important 
and lead to a much improved agreement of the theoretical and 
experimental data.

While published photoemission studies on $ {\rm MoO_2} $ focused
on the angle-integrated analysis of the occupied bands,
\cite{beatham78,werfel83,prakash08a,prakash08b} a systematic
investigation of the band dispersions and the Fermi body by
angle-resolved photoemission spectroscopy seems to be still missing.
In addition, existing magneto-transport measurements addressed only
fractions of the Fermi surface.
\cite{volskii71,volskii73,volskii75,volskii79,teplinskii80,klimm97}
It is the purpose of the present work to make up for the apparent
lack in experimental data and to present a comprehensive study of
the electronic structure of $ {\rm MoO_2} $.

An important difference between dHvA and ARPES is that the former
requires very low temperatures whereas the latter does not. This
is crucial for the early transition-metal dioxides, some of which
have a metal-insulator transition and thus do not show a Fermi
surface at low temperatures. In the present case we have the opportunity 
to compare ARPES and dHvA because $ {\rm MoO_2} $ is metallic at 
low $ T $. It turns out that the ARPES and dHvA results agree well, 
implying that ARPES will be useful to study the Fermi surfaces of 
the $ d^1 $ members $ {\rm VO_2} $ and $ {\rm NbO_2} $ as well.

The paper is organized as follows: While Sec.\ \ref{theory} gives
an overview on the computational method and the theoretical
results, Sec.\ \ref{experiment} turns to a description of the
ARPES and dHvA measurements and their results. The theoretical
and experimental findings are subject of a comparative discussion
given in Sec.\ \ref{discussion} and Sec.\ \ref{summary}
summarizes our results.

\section{Electronic structure calculations}
\label{theory}

\subsection{Computational Method}

The calculations are based on density-functional theory and
the local density approximation. \cite{hohenberg,kohnsham}
They were performed using
the scalar-relativistic implementation of the augmented spherical
wave method (see Refs.\ \onlinecite{aswrev,aswbook} and
references therein). In the ASW method, the wave function is
expanded in atom-centered augmented spherical waves, which are
Hankel functions and numerical solutions of Schr\"odinger's
equation, respectively, outside and inside the so-called
augmentation spheres. In order to optimize the basis set,
additional augmented spherical waves were placed at carefully
selected interstitial sites. The choice of these sites as well as
the augmentation radii were automatically determined using the
sphere-geometry optimization algorithm. \cite{sgo} Self
consistency was achieved by a highly efficient algorithm for
convergence acceleration. \cite{mixpap} The Brillouin zone
integrations were performed using the linear tetrahedron method
with up to 518 {\bf k}-points within the irreducible wedge.
\cite{bloechl94,aswbook}

In the present work, we used a new full-potential version of the
ASW method, which was implemented only very recently. \cite{fpasw}
In this version, the electron density and related quantities are
given by a spherical harmonics expansion inside non-overlapping muffin-tin
spheres. In the remaining interstitial region, a representation in
terms of atom-centered Hankel functions is used. \cite{msm88}
However, in contrast to previous related implementations, it is
sufficient to work without having to employ a so-called
multiple-$ \kappa $ basis set, permitting a very high computational
speed of the resulting scheme.

\subsection{Results}

The electronic properties of $ {\rm MoO_2} $ in the monoclinic as
well as in the above-mentioned artificial rutile structure have
already been discussed in much detail in our previous work.
\cite{habil,eyert00} For this reason, we will in the following
concentrate on the changes coming with the consideration of the
full potential rather than its muffin-tin approximation. In
particular, we will investigate the changes of the electronic
states near the Fermi energy.

Furthermore, since a very detailed description of the monoclinic
crystal structure and its relation to the rutile structure has
been given in our previous work, we will not enter a renewed
discussion here. However, we refer the reader to our previous
publication especially concerning the definition of the local
coordinate systems to be used for the analysis of the electronic
states. \cite{habil,eyert00}

\begin{figure}[htb]
\centering
\includegraphics[width=\columnwidth,clip]{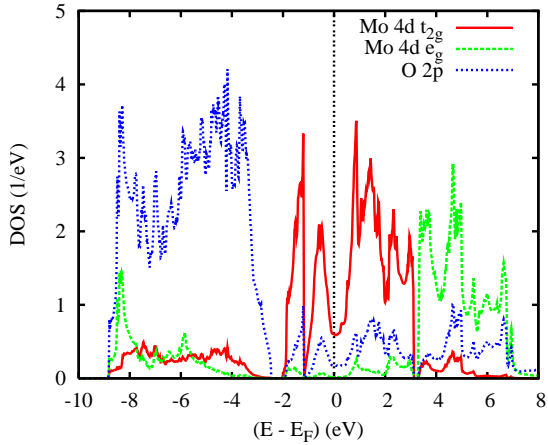}
\caption{(Color online) Partial densities of states (DOS).}
\label{theofig1}
\end{figure}

\begin{figure}[htb]
\centering
\includegraphics[width=\columnwidth,clip]{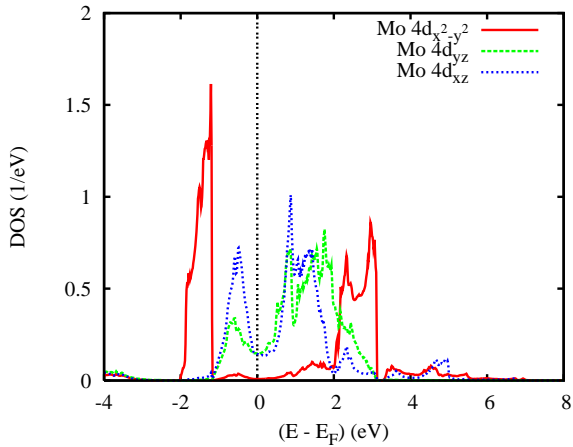}
\caption{(Color online) Partial Mo $ 4d $ $ t_{2g} $ DOS. Selection
         of orbitals is relative to the local rotated reference frame,
         see \protect Ref.\ \onlinecite{habil,eyert00}.}
\label{theofig2}
\end{figure}

While a first account of the monoclinic crystal structure was given
by Magn\'{e}li and Andersson, \cite{magneli55}, our calculations
are based on the refined data by Brandt and Skapski. \cite{brandt67}
Hence, in contrast to our previous work, \cite{habil,eyert00} we will
exclusively deal with the measured monoclinic structure
and not discuss our calculations for the related but artificial
rutile structure. In addition, since there are no indications for
the formation of magnetic moments in $ {\rm MoO_2} $, spin degeneracy
was enforced in all calculations. The resulting partial densities
of states (DOS) are shown in Fig.\ \ref{theofig1}.
They look very similar to those presented previously. 
\cite{habil,eyert00} In particular, we recognize the 6.5\,eV wide 
O $ 2p $ bands well below the Fermi energy $ E_{\rm F} $. In contrast, 
the Mo $ 4d $ $ t_{2g} $ and $ e_g $ manifolds are found around and 
well above $ E_{\rm F} $. Worth mentioning is the pronounced dip of 
the $ t_{2g} $ partial DOS at $ E_{\rm F} $. It results from the 
Peierls-like instability of the $ d_{x^2-y^2} $ states, which usually
are designated as the $ d_{\parallel} $ states and which mediate the
metal-metal overlap along the Mo atom chains parallel to the 
rutile $ {\bf c }$-axis. The splitting of these bands into bonding 
and antibonding branches as resulting from this Peierls-like 
instability is clearly visible in Fig.\ \ref{theofig2},
which displays the three partial $ t_{2g} $ DOS. In representing 
these partial DOS we have used the same local coordinate system as 
in our previous work with the local $ x $-axis parallel to the 
rutile $ {\bf c }$-axis and the local $ z $-axes pointing 
alternately along the $ [110] $ and $ [1\bar{1}0] $ direction.
\cite{habil,eyert00} In Fig.\ \ref{theofig2}, we observe the strong
splitting of the $ d_{x^2-y^2} $ states, which do no longer 
contribute to the metallic conductivity. Furthermore, due to the 
splitting, the low lying, bonding branch is found at the lower 
edge of the $ t_{2g} $ group of bands and nearly separated from 
the higher lying $ d_{xz} $ and $ d_{yz} $ bands. The latter are 
commonly designated as the $ \pi^{\ast} $ bands.

\begin{figure}[htb]
\centering
\includegraphics[width=0.5\columnwidth,clip]{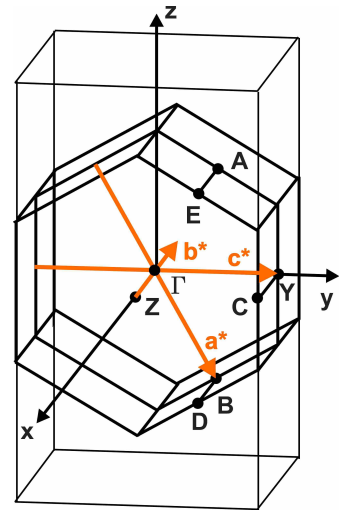}
\caption{(Color online) First Brillouin zone of the simple monoclinic
         lattice.}
\label{theofig3}
\end{figure}

\begin{figure}[htb]
\centering
\includegraphics[width=\columnwidth,clip]{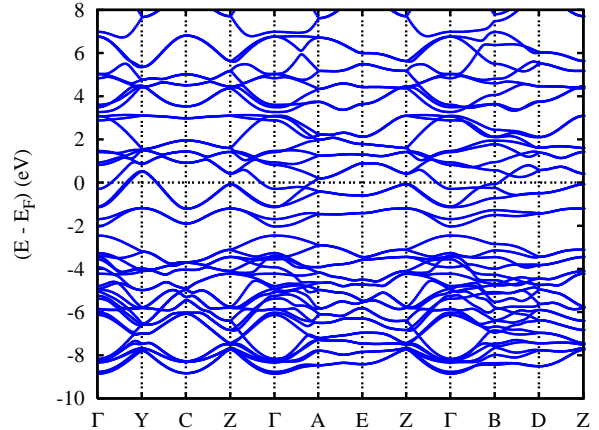}
\caption{(Color online) Electronic bands.}
\label{theofig4}
\end{figure}

\begin{figure}[htb]
\centering
\includegraphics[width=\columnwidth,clip]{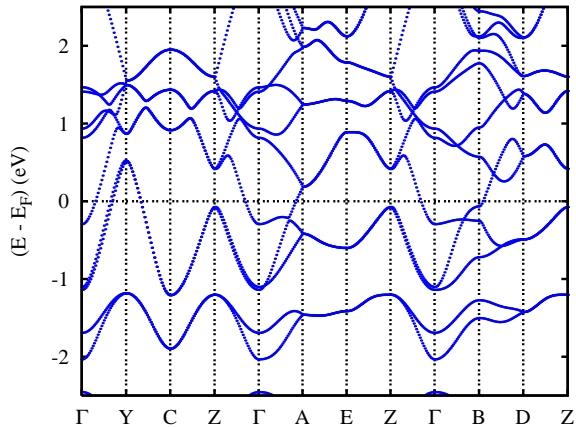}
\caption{(Color online) Near-$ E_{\rm F} $ electronic bands on an
         expanded energy scale.}
\label{theofig5}
\end{figure}

The electronic bands along selected lines within the first
Brillouin zone, Fig.\ \ref{theofig3},
are displayed in Fig.\ \ref{theofig4}.
With the interpretation of the partial DOS at hand, we easily
recognize the O $ 2p $ as well as the Mo $ 4d $ $ t_{2g} $ and $ e_g $
states in the energy intervals from $ -9 $ to $ -2.5 $\,eV,
$ -2 $ to $ +3 $\,eV, and $ +3 $ to $ +7 $\,eV, respectively. To
be more detailed, we find a two bands between $ -2 $ and $ -1 $\,eV,
which is separated from all other bands. This split-off doublet is
more easily identified in Fig.\ \ref{theofig5},
where we show the near-$ E_{\rm F} $ bands on an expanded energy scale.
According to the previous analysis, the split-off doublet is just the
bonding $ d_{\parallel} $ band. The one-dimensional behaviour of the
$ d_{\parallel} $ bands is contrasted by the rather isotropic dispersion
of the $ \pi^{\ast} $ bands. As already pointed out in our previous
work, hybridization between both types of bands is rather small. In
contrast coupling of these states is only via charge conservation.
For this reason, we proposed to interpret the splitting of the former
in the course of the structural distortions as an embedded Peierls-like
instability. \cite{habil,eyert00}

\begin{figure*}[htb]
\centering
\includegraphics[width=14cm,clip]{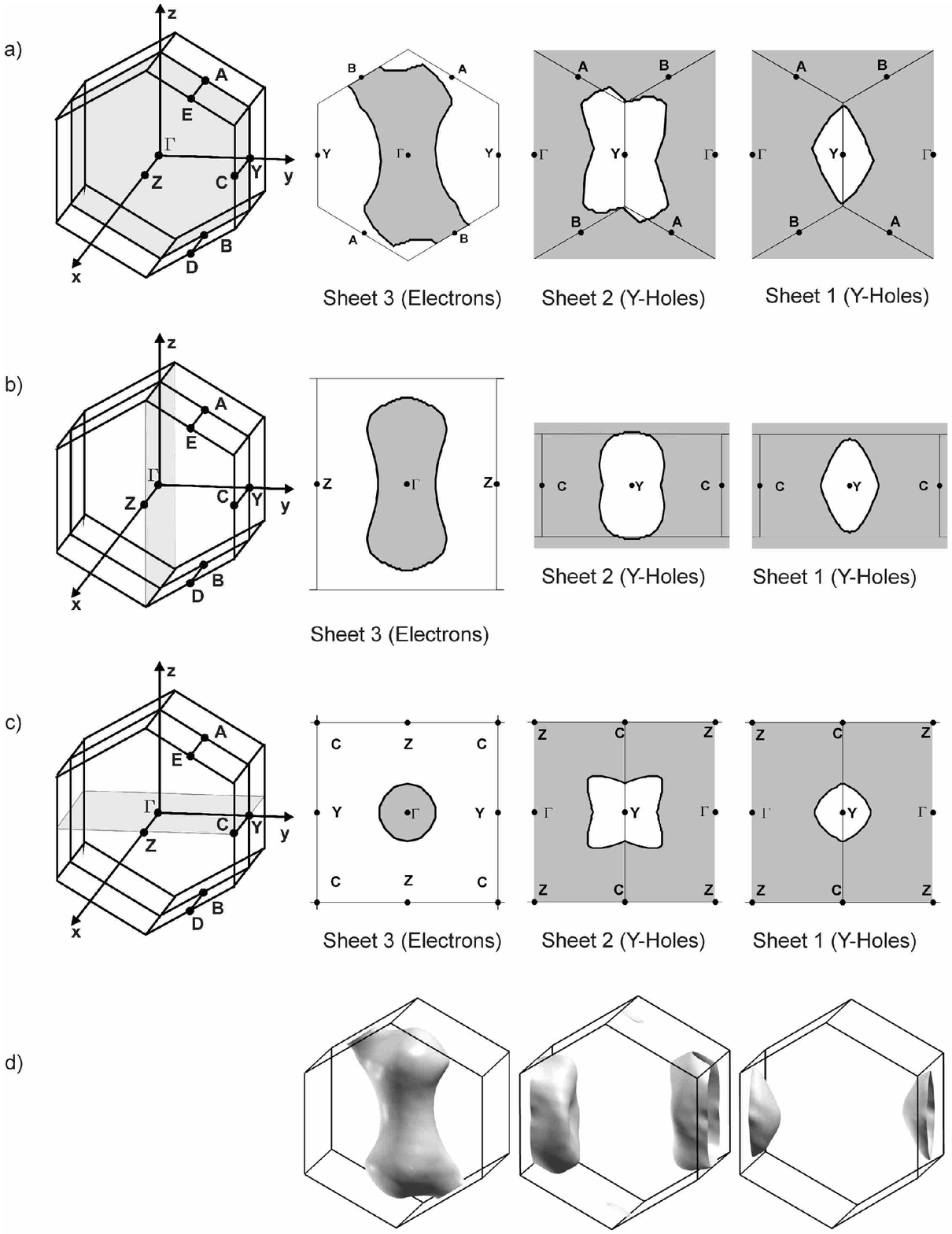}
\caption{Cuts through the Brillouin zone (left column) and the Fermi body
         a) in the ${\bf ac}$-plane,
         b) in the ${\bf ab}$-plane, and
         c) perpendicular to ${\bf a}$ as well as
         d) 3-dimensional view of the three Fermi surface sheets
         (generated using XCrysDen \protect \cite{kokalj03}).}
\label{theofig6}
\end{figure*}

Concentrating especially on the Fermi body of monoclinic $ {\rm
MoO_2} $, we complement Fig.\ \ref{theofig5} by Fig.\ \ref{theofig6},
which displays cuts through the Fermi body in the ${\bf ac}$-plane,
the ${\bf ab}$-plane, and perpendicular to the ${\bf a}$-axis as
well as a three-dimensional view of the Fermi surface sheets.

The Fermi body is defined by the Fermi level crossings of the $ d_{xz} $
and $ d_{yz} $ bands. In Figs.\ \ref{theofig5} and \ref{theofig6}, we
identify three different Fermi surface sheets: two hole-like surfaces
near the Y-point (sheet 1 and 2), which occupy 9 and 2\,\%, respectively,
of the volume of the 1st Brillouin zone, as well as a peanut-shaped
electron-like surface centered around the $ \Gamma $-point and occupying
11\,\% of the Brillouin zone volume (sheet 3). The latter sheet has
two loopholes near the B-points, whereas it is closed along the direction
$ \Gamma $-A.

As already mentioned in the introduction, use of the exact
single-particle potential, i.e.\ of the so-called full potential,
instead of its muffin-tin or atomic sphere approximation leads to
important improvements. These become obvious from comparing Fig.\
\ref{theofig5} of the present work to Fig.\ 6 of Ref.\
\onlinecite{eyert00}. Apart from a broadening of the O $ 2p $ bands
we find distinct differences especially at the Z- and the B-point.
To be specific, those bands, which form the highest occupied states
at these points in the present calculation, were found just above
$ E_{\rm F} $ in the old calculation. They generated small hole-like
pockets of the Fermi surface centered around these points, which were
at variance with the experimental findings (see below). In contrast,
in the new calculation these bands are found below $ E_{\rm F} $ and
the hole-like pockets disappear.

Another important difference can be observed at the $ \Gamma $-point. 
The maximum binding energy of the highest occupied band of the new 
calculation ($ 0.30 $\,eV) agrees much better with experimental value 
of $ 0.21 $\,eV (see below) than the binding energy of the old 
calculation ($ 0.43 $\,eV). Thus, by using the full-potential ASW 
method, the agreement between theory and experiment is considerably 
improved.

\section{Experiments}
\label{experiment}

\subsection{Sample preparation and characterization}

$ {\rm MoO_2} $ single crystals were grown by chemical transport
using $ {\rm TeCl_4} $ as transport agent. \cite{paulin96} The
crystals exhibit specular surfaces of a size up to $ 3 \times 3
$\,$ {\rm mm^2} $. The surface orientation was determined by x-ray
diffraction. For the ARPES measurements the as-grown surfaces were
cleaned in situ by heating to $ 750^\circ $\,C for some minutes.
The quality and cleanness of the heated single-crystal surfaces
were controlled by low energy electron diffraction (LEED) and by
scanning tunnelling microscopy (STM). For the dHvA measurements
several rods of $ 3 $\,mm length and $ 1.5 $\,mm diameter were
prepared. The $ {\bf b} $-axis was the axis of rotation of the rods.
The residual resistance ratio was 97 for
$ {\bf j} \parallel {\bf a} $ and 144 for $ {\bf j} \parallel {\bf b} $.

\subsection{Angle resolved photoemission spectroscopy (ARPES)}

\begin{figure}[htb]
\centering
\includegraphics[width=\columnwidth,clip]{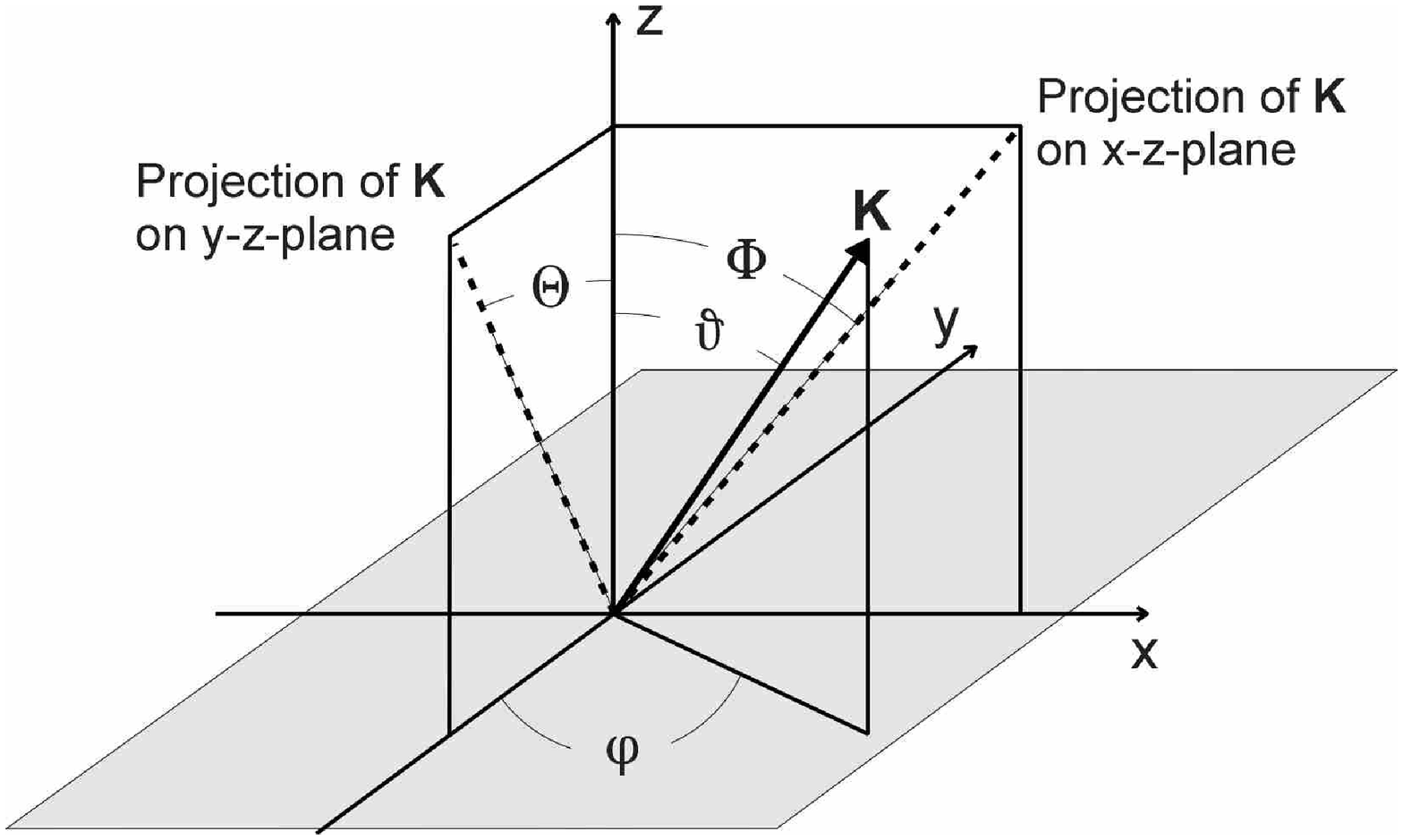}
\caption{Measurement geometry and definition of the  emission
         angles in an ARPES experiment. Here, $ {\bf K} $ is 
         the wave vector in vacuum.}
\label{expfig1}
\end{figure}

Experimentally, the electronic  structure $ E({\bf k}) $ was
investigated by ARPES measurements using the geometry and emission
angles defined in Fig.\ \ref{expfig1}.
Due to energy conservation the binding energy $ E_B $ of the
photoelectron is determined by
\begin{equation}
|E_{B}| = h\nu - E_{kin} - \phi, 
\label{Gl_Ekin}
\end{equation}
where $ h\nu $ is the photon energy and $ \phi $ the work 
function of the crystal. For a detailed description of the 
photoemission process we refer the reader to the literature.
\cite{cardona78,cardona79,huefner95,damascelli04}

On passing through the crystal surface, the component of the
electron wave vector parallel to the surface $ {\bf k}_{\parallel} $
remains unchanged with
\begin{equation}
   |{\bf k}_{\parallel}|
   =   \frac{1}{\hbar} \sqrt{2mE_{kin}} \sin \vartheta \;.
\label{Gl_k_par}
\end{equation}
The perpendicular component $ {\bf k}_{\perp} $ is, strictly spoken,
not conserved in the photoemission process but it is usually possible
to reconstruct its approximate value by including information about
the final states of the photoemission process. If one assumes a free 
electron-like final state dispersion cite{nielsen03} ,
$ k_{\perp} = |{\bf k}_{\perp}| $ can be calculated as 
\begin{equation}
{k}_{\perp}
  =   \frac{1}{ \hbar }
      \sqrt{ 2m ( E_{kin} cos^{2} \vartheta + |E_0|+ \phi ) } \;.
\label{Gl_ksenkr}
\end{equation}
Here, $ E_0 $ is the inner potential of the sample as referred to
$ E_{\rm F} $. While the work function $ \phi $ can be determined
experimentally, the inner potential $ E_0 $ is usually estimated
from an analysis of the measured band structure. As a consequence
of Eq.\ (\ref{Gl_ksenkr}), the set of all $ {\bf k}_{\parallel} $
vectors at a fixed photon energy describes a spherical surface in
$ {\bf k} $-space.

A Fermi vector $ {\bf k}_{\rm F} $ is defined by the Fermi level
crossing of a valence band in the corresponding direction of 
$ {\bf k} $-space. As ARPES permits a complete mapping of a solid's
three-dimensional band structure, the position of the Fermi level
crossings can be identified. A three-dimensional ARPES measurement
can be performed by varying the emission angles and $ h\nu $,
causing a variation of $ {\bf k}_{\parallel} $ and $ {\bf
k}_{\perp} $, respectively.

The ARPES measurements were done at the beamline SGM3 at the
synchrotron ASTRID in Aarhus, Denmark. The beamline provides light
in the energy range $ 12 $\,eV $ < h\nu < 130 $\,eV with an 
energy resolution of $ E / \Delta E = 15000 $. \cite{hoffmann04} 
The UHV chamber is equipped with an ARUPS10 electron analyzer 
(Vacuum Generators) with an angular resolution of $ 0.7^\circ $ 
and an energy resolution of $ 25 $\,meV. The analyzer is fixed on 
a computer controlled two-axis goniometer with an angular accuracy
up to $ 0.2^\circ $. All ARPES data were taken at a temperature of
$ 30 $\,K. For the description of the measurement geometry, instead 
of the polar emission angle $ \vartheta $ its projections on the 
$ xz $- and $ yz $-planes are used (see Fig.\ \ref{expfig1}). The 
resulting emission angles are called $ \Phi $ and $ \Theta $. 
The direction of normal emission was determined by a careful optical 
alignment of the highly reflective surface using a laser.

\subsection{ARPES Results}

\subsubsection{$ (100) $ surface}

\begin{figure}[htb]
\centering
\includegraphics[width=0.9\columnwidth,clip]{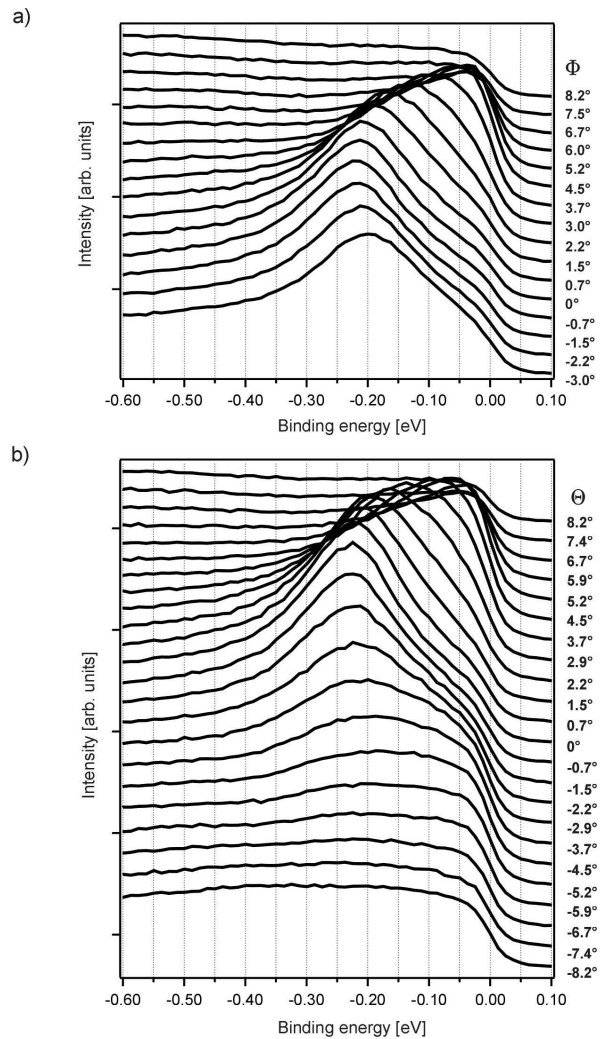}
\caption{ARPES measurements on the $ (100) $ surface under variation
         of the emission angle a) $ \Phi $ and b) $ \Theta $
         ($ h\nu=19 $\,eV).}
\label{expfig2}
\end{figure}

ARPES spectra taken on the $ (100) $ surface for variation of the
emission angles $ \Phi $ and $ \Theta $ are shown in Fig.\
\ref{expfig2}a and b,
respectively. At $ h\nu = 19 $\,eV, binding energies up to 
$ 0.6 $\,eV have been measured. In both graphs a single peak is
observed, which displays a nearly parabolic dispersion. At normal
emission ($ \Phi = 0 $ and $ \Theta = 0 $), it reaches a maximum
binding energy of $ 0.22 $\,eV. With increasing emission angles,
both $ \Phi $ and $ \Theta $, the peak approaches the Fermi level
and and crosses it. As will be shown later by the symmetry of the
Fermi surface, the $ \Gamma $-point is reached at $ h\nu = 19$~eV. Hence, 
the maximum binding energy of $ 0.22 $\,eV corresponds to the binding
energy at the $ \Gamma $-point. We can thus compare the occupied 
bandwidth determined here directly with the binding energy of the 
highest occupied LDA band at the $ \Gamma $-point which is $ 0.3 $\,eV.

Analyzing the nearly parabolic dispersion of the ARPES band,
effective masses of $ 0.876 m_e $ and $ 0.821 m_e $ can be found
for variation of the emissions angles $ \Phi $ and $ \Theta $,
respectively. This is in excellent agreement with Shubnikov-de
Haas measurements of Ref. \onlinecite{klimm97}, which provided for
comparable geometry an effective mass of $ 0.83 m_e $.

\subsubsection{Determination of the Fermi body}

The Fermi-level crossing of the highest occupied band, defining
the corresponding Fermi vector, can be identified by a strong
increase of photoemission intensity at $ E_{\rm F} $. Indeed,
ARPES can also provide a direct image of the Fermi surface when
the photoemission intensity in a narrow energy window around the
Fermi level is displayed as a function of emission angle or photon
energy (see Ref. \onlinecite{nielsen03} and references therein).

For a quantitative analysis of the ARPES spectra a constant
background as determined from the energy region of unoccupied
states has been subtracted. To correct intensity variations caused
by different emission angles, the spectra were normalized to this
background. The intensity dependence on $ h\nu $ is considered 
by normalization with the average total intensity of all spectra
taken at one certain photon energy. The work function has been
determined by VLEED measurements to $ \phi = 5.15 \pm 0.3 $\,eV.
\cite{moosburger05} The final electron states are supposed to be
free-electron like. The two following methods are used to extract
the Fermi vectors from the ARPES data:
\begin{enumerate}
\item The photoemission intensity is integrated over a $ 120 $\,meV
      wide energy region near $ E_{\rm F} $. The maxima of this
      integrated intensity $ I(\bf k) $ define the Fermi vectors.
\item Since the Fermi surface is determined by strong variations
      of the intensity distribution $ I(\bf k) $, the gradient
      $ |\nabla_{\bf k} I({\bf k})| $ could be likewise used to
      define the Fermi vectors. \cite{straub97,kipp99} However, this
      gradient method generally produces two Fermi contour lines.
      Whereas the contour line on the unoccupied side of the Fermi
      surface coincides with the real Fermi surface, the other line
      is an artefact of this technique. \cite{straub97}
\end{enumerate}
Both methods can be found in the literature. However, the gradient method
often results in a better agreement with theoretical predictions.
\cite{straub97}

\subsubsection{Fermi surface scans perpendicular to the direction
               $ {\bf a^{\ast}} $}

\begin{figure}[htb]
\centering
\includegraphics[width=\columnwidth,clip]{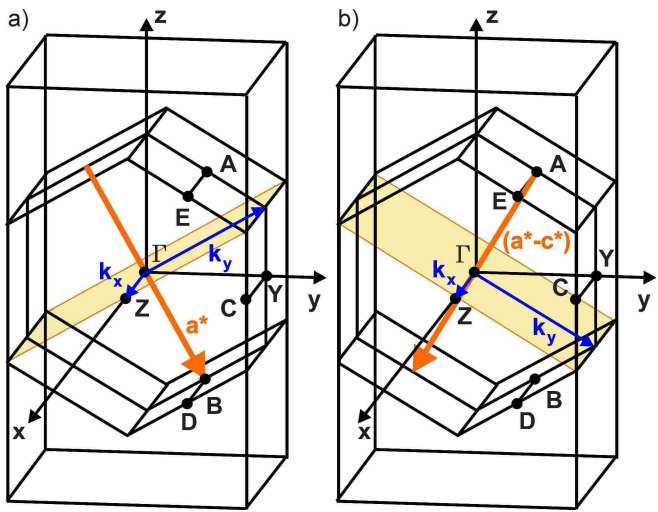}
\caption{(Color online) $ {\bf k} $-space cuts corresponding to
         $ {\bf k}_{\parallel} $ measurements in the
         a) $ (100) $ surface and
         b) $ (10\bar1)$ surface.
         The spherical curvature of the area in ARPES measurements
         is neglected.}
\label{expfig4}
\end{figure}

In the following, ARPES $ {\bf k}_{\parallel} $ scans on the $
(100) $ surface will be presented, which allow for a Fermi surface
mapping of planes perpendicular to the direction $ {\bf a^{\ast}} $.
Such Fermi surface maps can be performed by variation of both
emission angles at a fixed photon energy. They provide spherical cuts
through $ {\bf k} $-space. Nevertheless, for representation reasons,
flat projections of the spherical paths are shown. The theoretical
data are produced in complete analogy: spherical paths in
$ {\bf k} $-space are calculated, but flat projections are
shown. In order to simplify the discussion, we define a Cartesian
coordinate system within the plane of the projection by $ k_x $ and
$ k_y $, which are the components of the vector $ {\bf k}_{\parallel} $
parallel and perpendicular to the direction $ \Gamma $-Z, respectively.
This situation is sketched in Fig.\ \ref{expfig4}a.

\begin{figure}[htb]
\centering
\includegraphics[width=\columnwidth,clip]{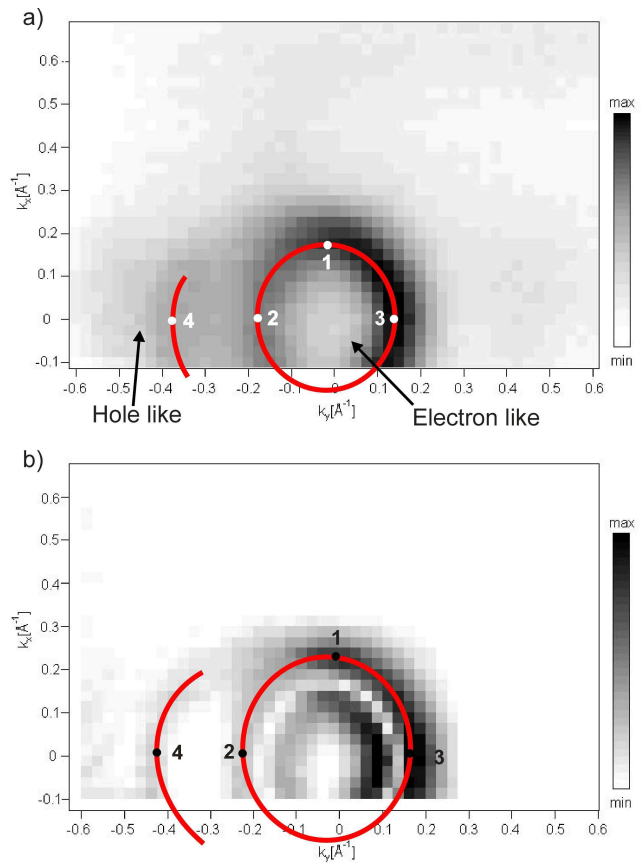}
\caption{(Color online) $ {\bf k}_{\parallel} $ mapping of the
         a) Photoemission intensity $ I({\bf k}) $ at $ E_{\rm F} $, and
         b) gradient of the photoemission intensity
         $ |\nabla_{\bf k}I({\bf k})| $ at $ E_{\rm F} $,
         both resulting from an ARPES measurement on the $ (100) $
         surface taken at a photon energy of $ h\nu = 19 $\,eV.
         Solid (red) lines point to structures of maximal intensity.
         Points 1 to 4 mark the Fermi vectors in direction of $ k_x $
         and $ k_y $.}
\label{expfig5}
\end{figure}

In Fig.\ \ref{expfig5}a,
an ARPES $ {\bf k}_{\parallel} $ scan through the Fermi body of 
$ {\rm MoO_2} $ is displayed. The measurement has been performed 
at a photon energy of $ h\nu = 19$\,eV. The photoemission intensity
at $ E_{\rm F} $ is plotted as a function of $ k_x $ and $ k_y $
using the intensity method (method 1). Two different Fermi surface
structures, characterized by an enhanced photoemission intensity,
can be identified, which are marked by solid (red) lines. Most
prominent is the ellipsoidal structure, which is located around
the origin and characterized by the points 1 to 3 with components
$ k_{xF1} = 0.175 $\,\AA$^{-1}$, $ k_{yF2} = -0.175 $\,\AA$^{-1}$,
and $ k_{yF3} = 0.150  $\,\AA$^{-1}$, respectively. In addition to
this electron-like Fermi surface a much less pronounced hole-like
Fermi surface contour is observed, albeit for negative values of $
k_y $ only. It comprises point 4 with component $ k_{yF4} = -0.375
$\,\AA$^{-1}$ and separates electron-like states with $ k_y >
k_{yF4} $ from hole-like states in the outer region $ k_y <
k_{yF4} $.

Figure \ref{expfig5}b shows the photoemission intensity gradient
$ |\nabla_{\bf k} I({\bf k})| $ as a function of $ k_x $ and $ k_y $
(method 2), based on the data shown in Fig.\ \ref{expfig5}a.
In Fig.\ \ref{expfig5}b, the above-mentioned double structure of
the gradient method is nicely visible for the electron-like Fermi
surface centered around the origin. The real Fermi surface contours,
i.e.\ those not caused by the artefact of the gradient method, are
marked by solid (red) lines. The extent of this Fermi surface
amounts to $ k_{xF1} =  0.225 $\,\AA$^{-1}$,
$ k_{yF2} = -0.213 $\,\AA$^{-1}$, and $ k_{yF3} = 0.188 $\,\AA$^{-1}$,
respectively. The hole-like Fermi surface contour for negative values
of $ k_y $ is also observed in Fig.\ \ref{expfig5}b. However, the
photoemission gradient in this region is very small and, hence, the
expected double Fermi surface structure can not be detected. Only the
real Fermi surface contour is visible. The extent of the hole-like
Fermi surface amounts to $ k_{yF4} = -0.438 $\,\AA$^{-1}$.
In general, the error of both methods of Fermi surface
determination is defined by the $ {\bf k} $-resolution of the ARPES
experiment and amounts to $ |\Delta {\bf k}| \approx 0.025 $\,\AA$^{-1}$.

\begin{figure}[htb]
\centering
\includegraphics[width=\columnwidth,clip]{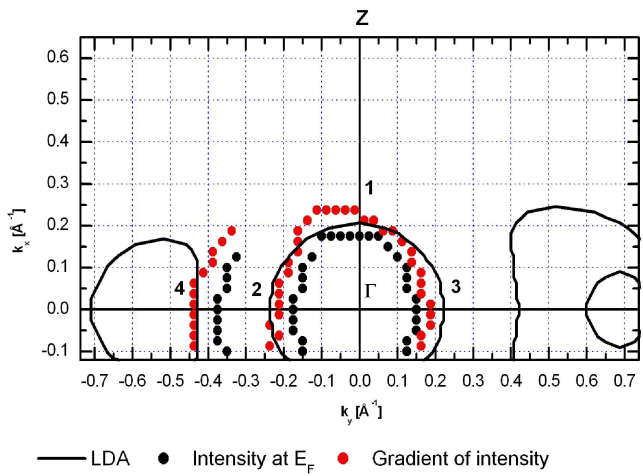}
\caption{(Color online) Cut through the Fermi body perpendicular
         to $ {\bf a}^{\ast} $: ARPES and LDA Fermi surface contours.}
\label{expfig7}
\end{figure}

In Fig.\ \ref{expfig7},
the experimental ARPES Fermi surface contours as resulting from
both the use of the intensity at $ E_{\rm F} $ and the gradient of
the intensity are compared to the contours resulting from the LDA
calculations. As mentioned above, the LDA calculations are performed
on spherical paths, so as to account for the spherical curvature of
the experimental plane. However, in the figures, the flat projection
on the plane perpendicular to $ {\bf a}^{\ast} $ is shown. Only the
$ {\bf k} $-space region accessible by the ARPES apparatus is included.

According to Fig.\ \ref{expfig7}, the LDA calculations predict an
electron-like, elliptic Fermi surface around the $ \Gamma $-point.
Note that, as will be shown later, $ k_{\perp} $ reaches the 
$ \Gamma $-point right at $ h\nu \approx 19 $\,eV. As a consequence,
the $ {\bf k}_{\parallel} $ mappings shown in Figs.\ \ref{expfig5}a 
and b as well as in Fig.\ \ref{expfig7} cover a plane perpendicular 
to $ {\bf a}^{\ast} $, which includes the $ \Gamma $-point (see Fig.\ 
\ref{expfig4}a). The theoretical findings are confirmed by the ARPES 
measurements. However, as compared to the LDA results, the Fermi 
surface contours determined by the intensity method are shifted 
towards the occupied states.  Hence, the ARPES Fermi surface is smaller 
than the calculated one with the deviations of the measured Fermi vectors 
ranging from $ -5 $ to $ -32 $\,\%. Here and in the following all 
deviations are given with respect to the calculated values. 
In contrast, the gradient method produces smaller shifts in both 
directions, away from as well as towards the occupied states. In this 
case, the deviations of the measured from the calculated values range 
from $ -15 $ to $ 20 $\,\%.

Perpendicular to the $ \Gamma $-Z direction, cuts through two 
hole-like Fermi surface sheets centered around the Y-point are 
predicted by theory (see also Fig.\ \ref{theofig6}a). However, 
owing to the spherical curvature of the measurement plane, observation 
of both hole-like Fermi surface contours is expected only for positive 
values of $ k_y $. In contrast, for $ k_y < 0 $ only one of these 
contours is cut by the measurement plane, which is indeed observed
in the ARPES experiments. In the region of negative $ k_y $ its inner 
surface can be seen in Fig.\ \ \ref{expfig7}, while both the outer 
part in $ k_y < 0 $ and the two hole-like structures for $ k_y > 0 $ 
are not detected. Compared to the LDA calculations, also the hole-like 
ARPES Fermi surface contour shows a slight shift, which, for the 
results obtained by the intensity method (method 1), is directed 
towards the occupied states. The corresponding ARPES Fermi surface 
is thus bigger than the calculated one, and the experimental deviations
from the theoretical values amount to $ 13 $ to $ 24 $\,\%. For the 
gradient method the ARPES contours generally agree well, but 
significant shifts with deviations between $ -2 $\,\% and $ 13 $\,\% 
are observed near some fractions of the Fermi surface. No systematic 
trend for the direction of these shifts can be discerned.

The experimental and theoretical values of the Fermi vectors $ k_{xF} $
and $ k_{yF} $ in the plane perpendicular to $ {\bf a}^{\ast} $ through
the $ \Gamma $-point as well as the deviations of the theoretical values
from the experimental ones as arising from the gradient method are
summarized in Tab.\ \ref{exptab1}.
\begin{table}[htb]
\begin{center}
\begin{tabular}{|c||c|c|c|c|}
\hline
                       & $ k_{x,F} $ & \multicolumn{2}{c|}{$ k_{y,F} $}
                                                            & $ k_{y,F} $ \\
                       & point~1   & point~2    & point~3   & point~4    \\
\hline
                       & \multicolumn{3}{c|}{ellipse} &  \\
\hline
\hline
$ I( k ) $             & $ 0.175 $   & $ -0.175 $  & $ 0.150 $   & $ -0.375 $ \\
\hline
$ |\nabla_{k}I({k})| $ & $ 0.225 $   & $ -0.213 $  & $ 0.188 $   & $ -0.438 $ \\
\hline
LDA                    & $ 0.207 $   & $ -0.237 $  & $ 0.222 $   & $ -0.429 $ \\
\hline
Deviation              & $   9 $\,\% & $ -10 $\,\% & $ -15 $\,\% & $ -2 $\,\% \\
\hline
\end{tabular}
\end{center}
\caption{Magnitudes of Fermi vectors in inverse \AA. The experimental
         Fermi vectors are taken from $ {\bf k}_{\parallel} $ scans
         on the $ (100) $ surface ($ h\nu = 19 $\,eV) and analyzed
         using both the intensity maximum and its gradient. The last
         lines gives the respective deviations of the experimental
         result as based on the gradient method from those arising
         from the LDA calculations.
         The plane perpendicular to $ {\bf a}^{\ast} $ through the
         $ \Gamma $-point has been analyzed. The points 1 to 4
         are marked in \protect Fig.\ \ref{expfig7}.}
\label{exptab1}
\end{table}

\subsubsection{Fermi surface scans in the
               $ {\bf a}^\ast {\bf c}^\ast $-plane}

A variation of $ {\bf k}_{\perp} $ can be achieved by varying the
photon energy (see Eq.\ (\ref{Gl_ksenkr})). In the so-called angle-energy 
scans one of the emission angles as well as the photon energy 
$ h\nu $ are varied while the respective other emission angle is 
held constant. From this, Fermi surface scans as a function of 
$ {\bf k}_{\perp} $ and one of the parallel wave vector components 
($ k_x $ or $ k_y $) can be obtained. In particular, angle-energy 
scans on the $ (100) $ surface under variation of $ h\nu $ and the 
emission angle $ \Theta $ (variation of $ k_y $) allow a Fermi 
surface mapping of the $ {\bf a}^\ast {\bf c}^\ast $-plane, which is 
spanned by the vectors $ {\bf a}^\ast $ and $ {\bf c}^\ast $. With 
$ k_x =0 $ ($ \Phi = 0 $), a cut through the $ \Gamma $-point is 
achieved. In this situation, $ {\bf k}_{\perp} $ is parallel to the 
direction $ \Gamma $-B, whereas $ k_y $ is orientated perpendicular 
to both $ \Gamma $-B and $ \Gamma $-Z. The thus described cut through 
the Brillouin zone is sketched in the left column of Fig.\ \ref{theofig6}a.

\begin{figure}[htb]
\centering
\includegraphics[width=\columnwidth,clip]{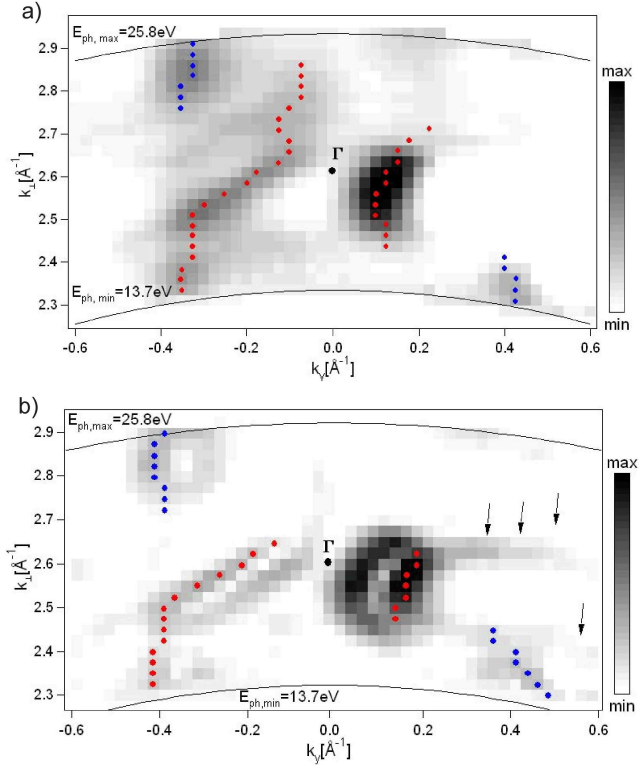}
\caption{(Color online) Angle-energy scan on the $ (100) $-surface:
         In dependence of $ k_y $ and $ k_{\perp} $
         a) the photoemission intensity at $ E_{\rm F} $, $ I({\bf k}) $, and
         b) the gradient of the photoemission intensity at $ E_{\rm F} $,
         $ |\nabla_{\bf k} I({\bf k})| $, is plotted. Artefacts are marked
         by arrows. (Light grey (red) points: electron-like Fermi surface
         contours, dark grey (blue) points: hole-like contours)}
\label{expfig9}
\end{figure}

The resulting ARPES Fermi surface is given in Fig.\ \ref{expfig9}a,
which shows the intensity $ I({\bf k}) $ as a function of $ k_y $ and
$ k_{\perp} $. Both, electron-like and hole-like Fermi surface contours
can be found, which are marked by light grey (red) and dark grey (blue)
points, respectively.

The inner potential, which is necessary for the calculation of 
$ k_{\perp} $, can be estimated by a comparison of the ARPES Fermi
surface contours to the LDA results. The LDA calculations give 
a peanut-shaped electron-like Fermi surface, which is symmetric with
respect to the $ \Gamma $-point (see Fig.\ \ref{theofig6}a). This
shape is reflected by the ARPES measurements, which likewise display 
the waist of a peanut of the electron-like Fermi surface. As already 
mentioned above, the center of this structure, which is just the 
$ \Gamma $-point, is reached at $ h\nu \approx 19 $\,eV. Using the 
parabolic free-electron like final state dispersion (third Brillouin 
zone), an inner potential of $ E_0 \approx -7 $\,eV is found. This 
value is comparable to the inner potential defined by the bottom of 
the valence band in LDA calculations and ARPES measurements 
($ E_{0, LDA} \approx -9 $\,eV).

Fig.\ \ref{expfig9}b displays the gradient of the intensity,
$ |\nabla_{\bf k}I({\bf k})| $, again as a function of $ k_y $ and
$ k_{\perp} $, based on the data shown in Fig.\ \ref{expfig9}a.
The typical twofold structure of real Fermi surface contours and the
technical artefacts can be observed. The real Fermi surface contours
are marked by points. Despite the normalization of the ARPES data,
there remain some variations of the photoemission intensity, which
are independent of the Fermi surface. In Fig.\ \ref{expfig9}b,
those artificial structures are marked by arrows. They will be
neglected in the subsequent discussion.

\subsubsection{Fermi surface scans perpendicular to the direction
               $ (\bf a^\ast-\bf c^\ast) $}

Fermi surface scans have been also performed on the $ (10 \bar 1) $
surface. If again the spherical curvature of the measurement plane
is neglected, the corresponding $ {\bf k}_{\parallel} $ scans on this
crystallographic orientation produce cuts through the Fermi body
perpendicular to the direction $ ({\bf a}^\ast - {\bf c}^\ast )$
(indicated in Fig.\ \ref{expfig4}b). In particular, while $ k_x $
is parallel to the direction $ \Gamma $-Z, $ k_y $ is perpendicular
to both the directions $ \Gamma $-Z and $ \Gamma $-A. This situation
is sketched in Fig.\ \ref{expfig4}b.

As a matter of fact, ARPES measurements on the $ {\rm MoO_2} (10\bar1)$
surface lead to less pronounced PES peaks than the measurements on the
$ (100) $ surface. This might be due to a higher roughness of the surface
or an enhanced density of defects.

\begin{figure}[htb]
\centering
\includegraphics[width=\columnwidth,clip]{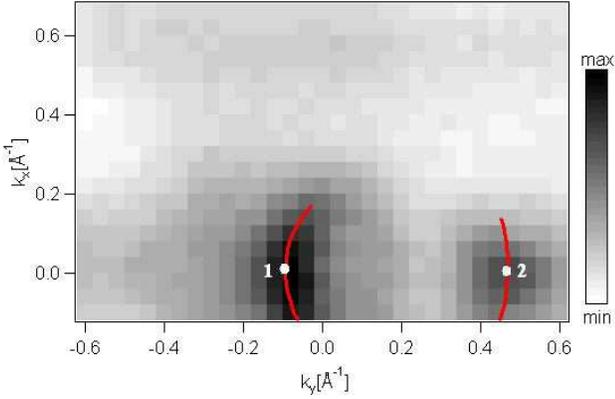}
\caption{(Color online) $ {\bf k}_{\parallel} $-mapping at the
         $ (10\bar1) $ surface at $ h\nu = 19 $\,eV. The
         photoemission intensity at $ E_{\rm F} $, $ I(\bf k) $ is
         plotted in dependence of $ k_x $ and $ k_y $. The solid
         (red) lines point up structures of maximal intensity.
         Points 1 and 2 mark the Fermi vectors in direction of
         $ k_x $ and $ k_y $.}
\label{expfig12}
\end{figure}

ARPES $ {\bf k}_{\parallel} $ scans on the $ {\rm MoO_2} (10\bar1) $ 
surface taken at $ h\nu = 19 $\,eV are displayed in Fig.\ \ref{expfig12},
where the intensity $ I(\bf k) $ is plotted as a function of $ k_x $ 
and $ k_y $. Near the origin, two separated structures of enhanced 
photoemission intensity can be observed, which in the direction $ k_y $ 
extent to $ k_{yF1} = -0.08 $\,\AA$^{-1}$ and 
$ k_{yF2} = 0.48 $\,\AA$^{-1}$, respectively. In contrast, in the
direction $ k_x $ no Fermi vector can be defined. The band crossing 
$ E_F $ in point 1 is occupied by electrons in the region
$ k_y > k_{yF1} $, the band responsible for point 2 is occupied by
electrons in the region $ k_y < k_{yF2} $. Unfortunately, the
photoemission peak creating these Fermi surface contours can not
be observed over the whole range of $ k_{yF1} < k_y < k_{yF2} $.
As a consequence, it is not clear, whether both crossing points
are generated by a single band or by two different bands. If they
are due to a single band, the region $ k_{yF1} < k_y < k_{yF2} $
must be electron-like. Then both Fermi vectors would be part of
the calculated peanut shaped, electron-like Fermi surface around
the $ \Gamma $-point (see Fig.\ \ref{theofig6}). In contrast, if
the crossing points were created by two different bands, only one
of the Fermi vectors could be part of the electron-like Fermi
surface.

Angle-energy scans on the $ (10 \bar 1) $ surface allow the
investigation of the $ {\bf a}^\ast {\bf c}^\ast $-plane (see
Fig.\ \ref{theofig6}a, left column), where $ {\bf k}_{\perp} $ is in the
direction of $ \Gamma $-A and $ k_y $ is orientated perpendicular to
both the directions $ \Gamma $-A and $ \Gamma $-Z. The corresponding
results will be included in the summary plot for the Fermi surface
contours in the $ {\bf a}^\ast {\bf c}^\ast $-plane to be presented
in the following section.

\subsubsection{Fermi surface in the $ {\bf a}^\ast {\bf c}^\ast $-plane:
               ARPES vs.\ LDA}

\begin{figure}[htb]
\centering
\includegraphics[width=0.6\columnwidth,clip]{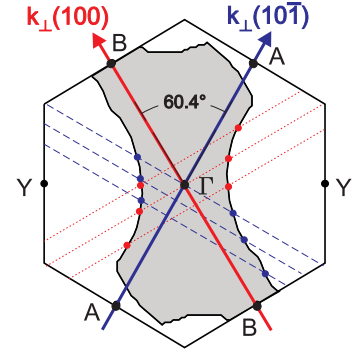}
\caption{(Color online) $ {\bf a}^\ast {\bf c}^\ast $-plane of the
         Brillouin zone and of the Fermi body: Angle-energy-scans
         on the $ (100) $ and $ (10\bar1) $ surfaces produce cuts
         perpendicular to the particular $ {\bf k}_{\perp} $. The
         end points of the corresponding Fermi vectors are marked
         by light grey (red) and dark grey (blue) points, respectively.}
\label{expfig13}
\end{figure}

So far, different ARPES mappings of the Fermi surface contours of
$ {\rm MoO_2} $ in the $ {\bf a}^\ast {\bf c}^\ast $-plane have
been described. Furthermore, detailed LDA calculations for this
plane have been done. In the following, all these data will be
combined into a single representation. To this end, the ARPES
results on the $ {\rm MoO_2} (100) $ surface and the theoretical
findings can be combined without further treatment. Yet, inclusion
of the measurements on the $ (10\bar1) $ surface is more complicated. 
In a combined plot of results from the (100) and (10-1) surface 
orientations, the respective $\Gamma$ points must
coincide. Furthermore, since the normal vectors of the $ (100) $
and $ (10\bar1) $ surface enclose an angle of $ 60.4^\circ $, the
perpendicular wave components $ {\bf k}_{\perp} $ of the two
surfaces also have to enclose this angle. The situation is
sketched in Fig.\ \ref{expfig13}.
For illustration the LDA Fermi surface and some exemplary cuts
through the Brillouin zone obtained by angle-energy-scans on the
two crystallographic surfaces have been included in this figure.
However, for each crystal surface there is an uncertainty of the
orientation of $ \pm 3 ^\circ $. Accordingly, for the angle enclosed
by the two surfaces an error of $ \pm 6 ^\circ $ is supposed.

\begin{figure*}[htb]
\centering
\includegraphics[width=14cm,clip]{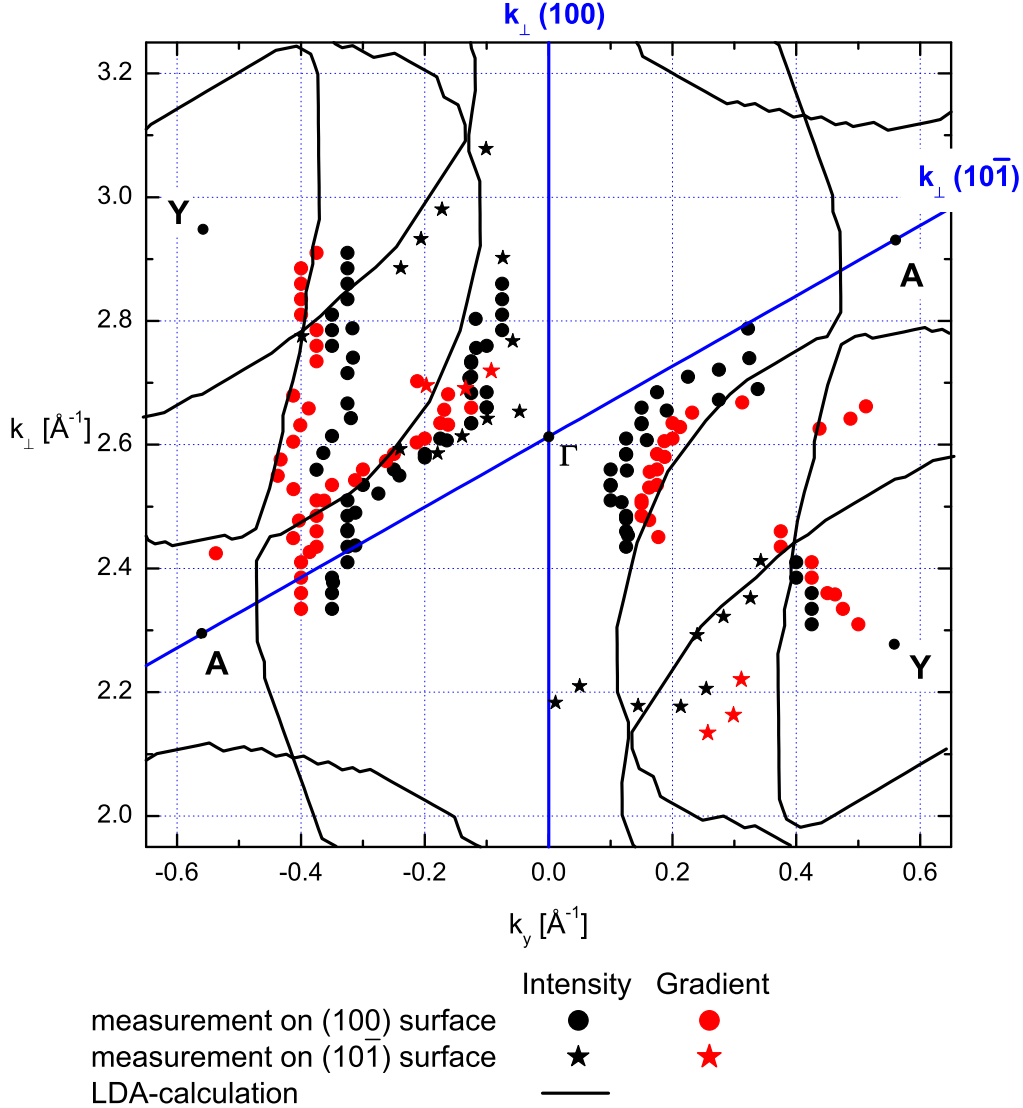}
\caption{(Color online) Cut through the Fermi body in the
         $ {\bf a}^\ast {\bf c}^\ast $-plane of $ {\rm MoO_2} $
         as indicated in \protect Fig.\ \ref{expfig13}:
         experimental Fermi surface contours from ARPES measurements
         on the $ (100) $ and $ (10\bar1) $ surfaces and theoretical
         Fermi surface contours from LDA calculations.}
\label{expfig14}
\end{figure*}

Of course, in the ideal case, the ARPES findings on the two different
crystallographic surfaces of $ {\rm MoO_2} $ should be consistent among
each other and they should be consistent with the LDA calculations.
Within the experimental errors, an adjustment of the inner potential
$ E_0 $, the work function $ \phi $ and the angle between the perpendicular
wave vectors is possible. The best agreement between all experimental and
theoretical findings is achieved for an inner potential of
$ E_0 = -8.5 $\,eV and a work function of $ \phi = 5 $\,eV for the
analysis of the data of the $ (10\bar1) $ surface and an angle of
$ 55.4 ^\circ $.  Using these values, we combine all experimental and
theoretical Fermi surface results on the
$ {\bf a}^\ast {\bf c}^\ast $-plane into Fig.\ \ref{expfig14}.
Obviously, the overall agreement between the experimental and theoretical 
findings is very good. Large parts of the theoretically predicted, 
peanut-shaped, electron-like Fermi surface around the $ \Gamma $-point 
are observed by the experiment. In particular, the waist of the 
``peanut'' is correctly confirmed by all ARPES measurements. Moreover, 
the two hole-like Fermi surface contours around the Y-point can also 
be observed in the ARPES measurements. Especially at 
$ k_y \approx -0.4 $\,\AA$^{-1}$, a considerable fraction of this 
Fermi surface is observed.

Nevertheless, a more detailed analysis between the experimental and
theoretical data is hampered by the dependence of the ARPES results
on the method of analysis. Use of the maximal intensity at the Fermi
level leads to Fermi surface contours, which compared to the
theoretical predictions are shifted towards the occupied states. As
a consequence, the corresponding electron-like Fermi surface is smaller,
the hole-like Fermi surface bigger than the LDA contours. In general,
the deviations are small but the error depends on the absolute size
of the Fermi surface and may reach $ -45 $\,\%
in a few directions. In contrast, analysis of the data with the help
of the gradient of the photoemission intensity delivers contours, which
agree much better with theory. However, again there exist a few single
directions, where the deviations from theory may go up to $ -30 $\,\%.
Nevertheless, as the comparison of different methods of ARPES data analysis
revealed, the gradient method delivers the best results. \cite{straub97,kipp99}
For this reason, only the results of the intensity gradient method
will be included in the following discussions.

Finally, in order to allow for a comparison with the de Haas-van Alphen
data to be presented in the subsequent section, the ARPES surface area
of the electron-like Fermi surface in the
$ {\bf b}^\ast  {\bf c}^\ast $-plane has also been determined and
compared to the LDA results. Under the assumption of an ellipsoidal
form the area can be calculated using the Fermi vectors in the directions
$ \Gamma $-Y (see Fig.\ \ref{expfig14}) and $\Gamma $-Z (see Fig.\
\ref{expfig7}). As a result, the deviation of the theoretical surface
area from the experimental one amounts to $ -7 $\,\% (see Tab.\
\ref{exptab2} below).

\subsection{De Haas-van Alphen (dHvA) oscillations}

The de Haas-van Alphen effect provides a well-established alternative
to study the Fermi surface of metallic systems. This quantum
mechanical effect manifests itself as an oscillation of the
magnetization/susceptibility as a function of an external magnetic
field. The basic ideas of magnetic oscillations in metals were first
described by Onsager, \cite{onsager1952} and later on detailed by
Shoenberg. \cite{shoenberg84}

The periodicity of the oscillations as function of the inverse of 
the magnetic field $ 1/|{\bf B}| $ can be associated with a frequency 
$ F $, which usually is given in units of Tesla $ [T] $. This frequency $ F $ 
is related to the extremal Fermi surface cross section $ A_{\rm Fermi} $ 
via the Onsager relation
\begin{equation}
F[T] = \frac{\hbar}{2 \pi e} A_{\rm Fermi}
\label{Gl_FrequenzdHvA}
\end{equation}
Variation of the direction of the magnetic field provides a good 
guess of the Fermi surface geometry. However, due to the ellipsoidal 
approximation, which is usually applied to derive the Fermi surface 
cut from the cross section $ A_{\rm Fermi} $, the shape of the Fermi 
surface may not be fully resolved. Hence, for more complex Fermi 
surfaces the evaluation of the data must be complemented, e.g., by 
LDA calculations. In passing we point to the well-known fact that, 
in contrast to the ARPES experiments discussed above, 
de Haas-van Alphen measurements allow to determine only the extremal 
cross sections of the Fermi body but do not have access to the 
complete band structure $ E(\bf k) $.

In a more quantitative treatment performed by Lifshitz and Kosevich
\cite{lifshitz56} and Shoenberg \cite{shoenberg84} the oscillating 
part of the magnetisation parallel to $ {\bf B} $ is given by the 
Lifshitz-Kosevich formula
\begin{equation}
\tilde M_{\|}
\propto \frac{F \sqrt{B}}{\sqrt{A''}}
        \sum_{p} R_T R_D R_S \frac{1}{p^{\frac{3}{2}}}
                 \sin{ \left[ 2 \pi p
                       \left( \frac{F}{B}-\frac{1}{2} \right)
                       \pm \frac{\pi}{4}
                       \right]
                     },
\label{Gl_Lifshitz}
\end{equation}
where $ B = |{\bf B}| $, $ A'' = \frac{d^2A}{dk^2} $, $ R_T $,
$ R_D $, and $ R_S $ are reduction coefficients, and the sum
includes all harmonics $ p $.

While a closer investigation of the measured signal reduction delivers
more details of the electronic structure, in this paper only the
temperature dependent factor $ R_T $ is of further interest. A finite
temperature $ T $ causes a broadening of the Fermi energy, which is
reflected by the reduction coefficient
\begin{equation}
R_T = \frac{X}{ \sinh{X} } \;, \qquad
  X = \frac{2 \pi^2 p k_B T}{\hbar \omega_c}.
\label{Gl_RedCoef}
\end{equation}
Finally, via temperature dependent measurements the effective mass
$ m_c $ of charge carriers included in $\omega_c =\frac{eB}{m_c}$
can be determined.

For the dHvA measurements a $ 15 $\,T magnet system in combination
with a $ ^3 $He/$ ^4 $He dilution refrigerator and a rotator of
Oxford Instruments was used. In our setup the field modulation
technique was used. Thus the dHvA magnetisation is a function of
$ B_0+b_0 \cos(\omega_{mod} t) $ and was detected via a mutual
inductance bridge (astatic coil pair). The measurements were
performed at $ B_0 $ between $ 10 $ and $ 15 $\,T with a field
variation rate of $ 15 $\,mT/min. The amplitude of the excitation
field $ b_0 $ was a few mT with a modulation frequency of
$ \omega_{mod} = 19 $\,Hz. With these values, heating of the
specimen by eddy currents can nearly be neglected and the
temperature was between $ 15 $ and $ 25 $\,mK.

\subsection{Results (dHvA)}

\begin{figure}[htb]
\centering
\includegraphics[width=\columnwidth,clip]{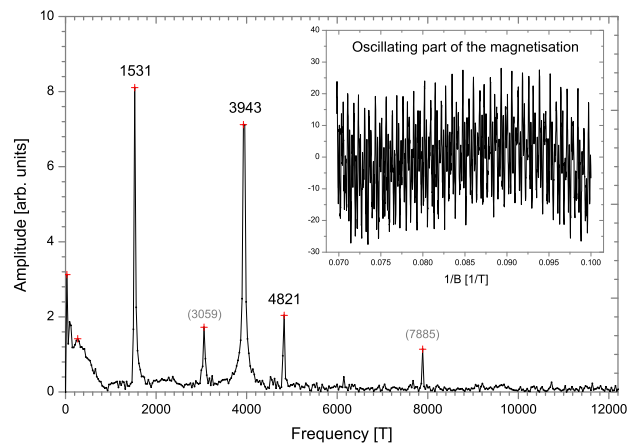}
\caption{(Color online) dHvA frequencies of extremal cross sections of
         the Fermi body in one direction. All extremal cross sections
         are captured simultaneously (1531\,T, 3943\,T, 4821\,T),
         higher harmonics are given in brackets (3059\,T, 7885\,T).
         The inset shows the underlying oscillating part of the
         magnetisation as function of $1/B$ after subtraction of a
         third order polynomial.}
\label{dHvA_oscillation}
\end{figure}

In order to determine the Fermi body of $ {\rm MoO_2} $, the angular
dependence of the dHvA oscillations was measured. With $ {\bf b}^\ast $
as the axis of rotation (which is perpendicular to the sheet plane of
the following figures) the angle of the magnetic field was varied in
the $ {\bf a}^\ast {\bf c}^\ast $-plane in steps of $ 15 ^\circ $. Due
to point symmetry the measurements ran only from $ 345 ^\circ$ down to
$ 150 ^\circ$ (relative to the ${\bf a} $-axis). To separate the
oscillating part of the signal from the background, a third order
polynomial was subtracted from the 2048 points equidistant in $ 1/B $
(see inset Fig. \ref{dHvA_oscillation}). A subsequent Fast Fourier
Transformation yielded the frequency spectrum in units of Tesla
(see Fig. \ref{dHvA_oscillation}).
After elimination of higher harmonics and occasional isolated points
the angular dependence of the frequencies -- and the corresponding
mirror points -- was plotted in polar coordinates. Each point in the
plot represents an area of extremal cross section of the Fermi body.
1\,T corresponds to an area in $ {\bf k} $-space of
$9.55 \cdot 10^{-5}$\,\AA$^{-2}$. It must be pointed out that all
extremal cross sections perpendicular to $ \bf B $ were captured
simultaneously. This complicated the classification of different
Fermi surface sheets and of possibly existing multiple extremal
cross sections of one Fermi surface sheet.

In the $ {\bf a}^\ast {\bf c}^\ast $-plane of $ {\rm MoO_2} $,
three different Fermi surface sheets have been detected, which
will be treated in detail in the following three subsections.

\subsubsection{dHvA measurement of sheet 1 (Y-hole)}

\begin{figure}[htb]
\centering
\includegraphics[width=\columnwidth,clip]{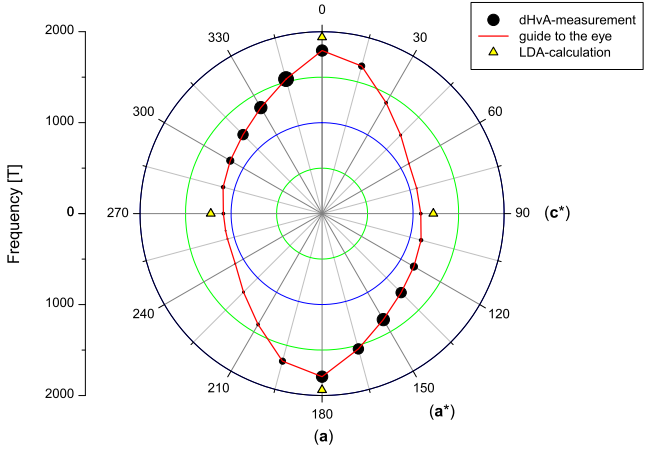}
\caption{(Color online) dHvA findings and LDA results of the hole-like
         Fermi surface around the Y-point (sheet 1) in the
         $ {\bf a}^\ast {\bf c}^\ast $-plane. The size of the
         experimental points corresponds to the intensity of the signal.}
\label{expfig15}
\end{figure}

In Fig.\ \ref{expfig15}, 
the angular dependence of the dHvA frequencies of one Fermi surface 
sheet in the $ {\bf a}^\ast {\bf c}^\ast $-plane is plotted. The 
correlation of the plotted points is well defined by the marginal 
change of frequency and the relative high intensity. The latter 
shows a maximum between $ 120 ^\circ $ and $ 180 ^\circ $ caused 
by a minimal curvature of the area 
($ \tilde M_{\|} \propto \frac{1}{\sqrt{A''}} $). The absence of 
appreciable intensity between $ 30 ^\circ $ and $ 75 ^\circ $ is 
an indication of deviations of the rotational symmetry referred to 
the $ {\bf a} $-direction. The deviations can be observed as a small 
bulge at an angle of $ 105 ^\circ $. The measured frequency amounts 
to 1126\,T, whereas the frequency at the corresponding angle (with 
respect to the rotational symmetry) of $ 255 ^\circ $ amounts only 
to 1076\,T. This variation of the magnitude of the bulge 
in different directions (regarding the whole three-dimensional Fermi 
surface) can explain 
differences of the signal intensities. The extension of the dHvA 
Fermi surface in the $ {\bf c}^\ast $-direction ($ \Gamma $-Y) 
amounts to 1084\,T, in direction perpendicular to $ {\bf c}^\ast $ 
to 1793\,T.

Shape and size of the experimental Fermi surface cross section
allow the assignment to one of the LDA Fermi surfaces (see Fig.\
\ref{theofig6}a). The hole-like Fermi surface structure centered
around the Y-point (sheet 1) gives a very similar shape with a
bulge at $ 100 ^\circ $. The size is likewise in good agreement
with the results of the calculations. A quantitative comparison
can be given for the $ {\bf c}^\ast $-direction and the direction
perpendicular to $ {\bf c}^\ast $. The discrepancy of dHvA extremal
surface areas to LDA calculations is $ -11 $\,\% in the
$ {\bf c}^\ast $-direction and $ -8 $\,\% perpendicular to
$ {\bf c}^\ast $. Hence, the experimental dHvA Fermi surfaces are
smaller than those theoretically predicted.

\subsubsection{dHvA measurement of sheet 2 (Y-hole)}

\begin{figure}[htb]
\centering
\includegraphics[width=\columnwidth,clip]{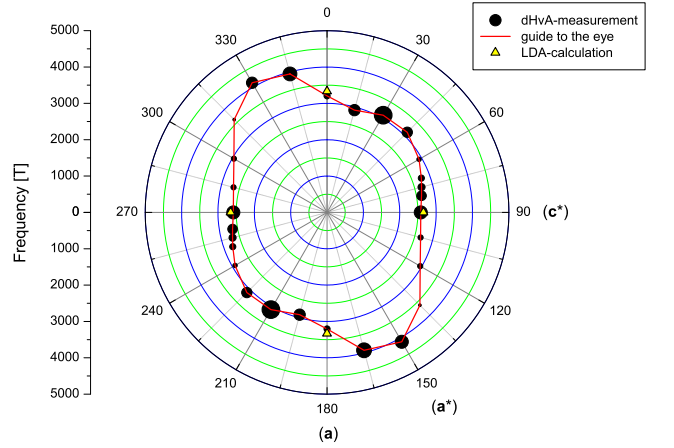}
\caption{(Color online) Same as Fig.\ \protect \ref{expfig15}, but for
          sheet 2 (Y-hole).}
\label{expfig16}
\end{figure}

In Fig.\ \ref{expfig16}, 
the angular dependence of the dHvA frequencies of a second Fermi 
surface sheet in the $ {\bf a}^\ast {\bf c}^\ast $-plane is shown.
The shape of the dHvA Fermi surface resembles the hole-like LDA 
Fermi surface around the Y-point, called sheet 2 (see Fig.\ 
\ref{theofig6}a). However, the contraction of the experimental dHvA 
Fermi surface in direction of $ {\bf c}^\ast $ seems to be smaller 
than theoretically predicted. Here the deficiencies of the ellipsoidal 
approximation appear. Perpendicular to the 
$ {\bf a}^\ast {\bf c}^\ast $-plane, in the 
$ {\bf b}^\ast {\bf c}^\ast $-plane, the Fermi surface sheet 2 can not 
be described by an 
ellipsoid. In this plane, the LDA calculations yield a star shaped 
Fermi surface with an area larger than that of an approximated 
ellipsoid (see Fig.\ \ref{theofig6}c). The deviations cause less 
contraction of the measured dHvA Fermi surface in 
$ {\bf c}^\ast $-direction as predicted by theory. The high intensity of the 
signal at $ 90 ^\circ$ indicates the continuation of the star 
shape of the $ {\bf b}^\ast {\bf c}^\ast $-plane over the whole 
side.

The measured frequencies amount to $ 2577 $\,T and $ 3205 $\,T,
parallel and perpendicular to the $ {\bf c}^\ast $-axis, respectively.
These results are $ -3 $\,\% and $ -4 $\,\% smaller than those
predicted by the LDA calculations.

\subsubsection{dHvA measurement of sheet 3 (Electrons)}

\begin{figure}[htb]
\centering
\includegraphics[width=\columnwidth,clip]{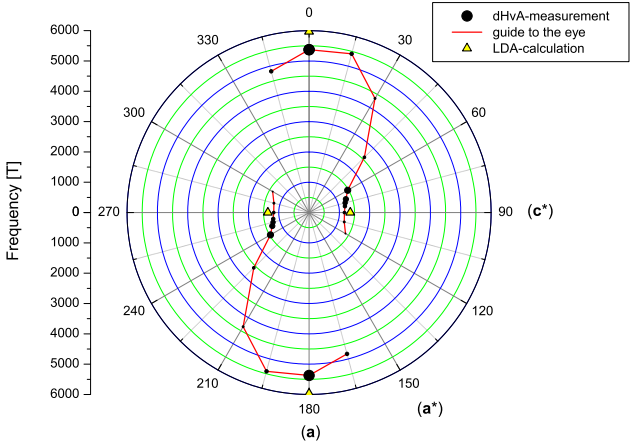}
\caption{(Color online) Same as Fig.\ \protect \ref{expfig15}, but for
          sheet 3 (electron pocket).}
\label{expfig17}
\end{figure}

Fig.\ \ref{expfig17},
shows the angular dependence of the dHvA findings of a third Fermi 
surface sheet in the $ {\bf a}^\ast {\bf c}^\ast $-plane. The shape 
is similar to the electron-like, peanut-shaped LDA Fermi surface 
sheet 3 (see Fig.\ \ref{theofig6}a). Although the approximate 
rotational symmetry with respect to the $ {\bf a} $-axis, which was 
obtained by the LDA calculations (see Fig.\ \ref{theofig6}c), 
simplifies the identification of the belonging points, there are 
large changes in frequency combined with weak intensity. The
frequency of the peanut waist in direction of $ {\bf c}^\ast $ 
is 1166\,T. The discrepancy to LDA calculations amounts to $ -14 $\,\%. 
In the $ {\bf a}$-direction a frequency of 5370\,T is found. Here 
the deviation from theory amounts to $ -10 $\,\%. Again, the experimental 
Fermi surface is smaller than the theoretical one.

The missing signal at $ 315 ^\circ $ and $ 330 ^\circ $ can be
explained by the open orbits caused by the neck of the peanut in
the direction $ \Gamma $-B.

\subsubsection{Fermi surface in the $ {\bf a}^\ast {\bf c}^\ast $-plane:
               dHvA vs.\ LDA}

In the following, the results of the dHvA measurements and the LDA
calculations in the $ {\bf a}^\ast {\bf c}^\ast $-plane will be
compared. In Tab.\ \ref{exptab2},
\begin{table}[htb]
\begin{center}
\begin{tabular}{|c|c||r|r|}
\hline
\multicolumn{2}{|c||}{}    & $ \| {\bf c}^\ast $ ($ 90^\circ $)
                           & $ \perp {\bf c}^\ast $ ($ 0 ^\circ $) \\
\hline
\hline
Sheet 1  & dHvA            & $ 1084 $\,T    & $ 1793 $\,T    \\
         & LDA             & $ 1222 $\,T    & $ 1939 $\,T    \\
         & (dHvA-LDA)/LDA  & $ -11 $\,\%    & $  -8 $\,\%    \\
\hline
Sheet 2  & dHvA            & $ 2577 $\,T    & $ 3205 $\,T    \\
         & LDA             & $ 2650 $\,T    & $ 3333 $\,T    \\
         & (dHvA-LDA)/LDA  & $  -3 $\,\%    & $  -4 $\,\%    \\
\hline
Sheet 3  & dHvA            & $ 1166 $\,T    & $ 5370 $\,T    \\
         & LDA             & $ 1358 $\,T    & $ 5968 $\,T    \\
         & (dHvA-LDA)/LDA  & $ -14 $\,\%    & $ -10 $\,\%    \\
         & (ARPES-LDA)/LDA & $  -7 $\,\%    &                \\
\hline
\end{tabular}
\end{center}
\caption{Areas of Fermi surface cross sections as resulting from the
         de Haas-van Alphen measurements and the calculations,
         respectively. The respective last lines for each sheet
         give the deviations of the experimental results from the
         calculated ones. For sheet 3, the corresponding deviation
         for the ARPES measurements has been added.}
\label{exptab2}
\end{table}
the dHvA surface areas and the LDA results parallel and perpendicular
to $ {\bf c}^\ast $ are summarized. All three theoretically predicted
Fermi surface sheets have been confirmed by the dHvA measurements.
Qualitatively, the shapes of the experimental and theoretical Fermi
surface contours agree well.

The quantitative comparison along selected high symmetry directions
shows that there are slight discrepancies between the dHvA and the
LDA results. All dHvA Fermi surface areas are smaller than calculated
by LDA. The discrepancies lie between $ -3 $\,\% and $ -14 $\,\%.
The peanut shaped, electron-like Fermi surface sheet 3 shows the
largest deviation to the LDA calculations.

Due to the underlying ellipsoidal approximation, the shape of the
dHvA Fermi surface sheets have to be treated with care. For some
directions the real Fermi surface topology clearly differs from the
assumed ellipsoidal shape. Especially, for sheet 2 ($ \| $ and
$ \perp $ to $ {\bf c}^\ast $) and for sheet 3 ($ \perp $ to
$ {\bf c}^\ast$) the LDA results indicate non-negligible deviations.

A comparison of our dHvA results to previously published magneto-transport 
data reveals a very good accordance.
\cite{volskii71,volskii73,volskii75,volskii79,teplinskii80} In
particular, the results for the Fermi surface sheet 1 agree nearly
perfectly, only differences of less then 1.5\,\% were found.
In contrast, only little has been published for sheets 2 and 3.
Yet, our results for these Fermi surface sheets, while being much
more comprehensive, agree very well with the existing data, the
differences being less than 6.5\,\%.

\subsubsection{Effective masses}

\begin{figure}[htb]
\centering
\includegraphics[width=\columnwidth,clip]{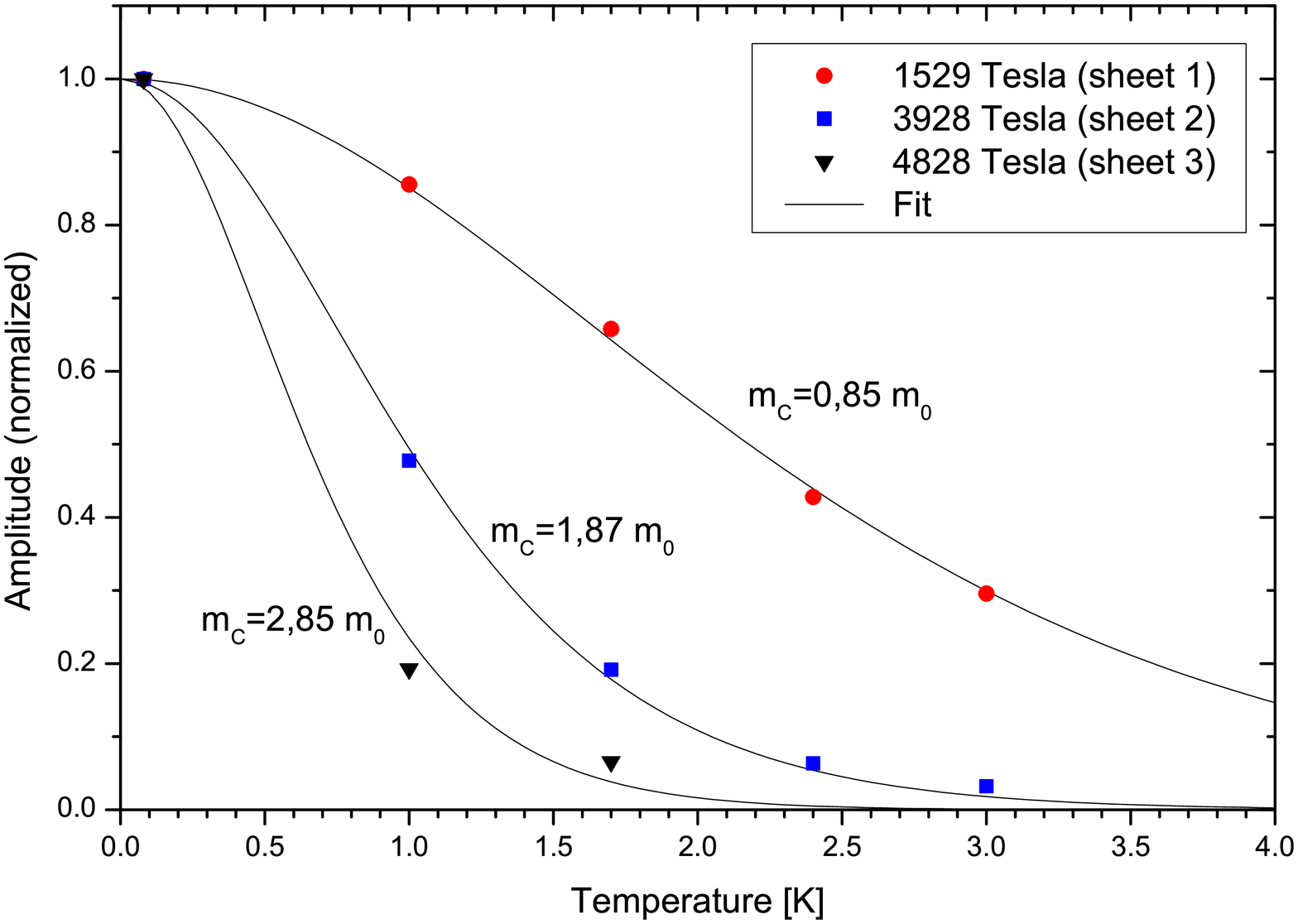}
\caption{(Color online) Temperature dependent measurement of the oscillation
         amplitudes for the three detected Fermi surface sheets and the
         corresponding effective masses using \protect Eq.\ (\ref{Gl_RedCoef})
         as a fit. The angle between $ {\bf B} $ and $ {\bf a} $
         amounts to $ 75 ^\circ $.}
\label{expfig18}
\end{figure}

Temperature dependent measurements of the magnetic oscillations 
allow the determination of effective masses of the charge carriers 
via Eq.\ (\ref{Gl_RedCoef}). An increase in temperature causes a 
reduction of the dHvA signal. For a single orientation of the magnetic 
field the effective masses of the three observed bands have been 
analyzed. The angle between $ {\bf B} $ and $ {\bf a} $ amounts to 
$ 75 ^\circ $. The results are displayed in Fig.\ \ref{expfig18}. 
The temperature has been varied between $ 0.1 $\,K and $ 3 $\,K. 
The oscillation amplitudes (normalized to one at $ T = 0.1 $\,K) as 
a function of the temperature are plotted. A fit of the data using 
Eq.\ (\ref{Gl_RedCoef}) yields the effective masses $ m_c $.

The effective masses of the three Fermi surface sheets differ
clearly. Sheet 1 shows a slightly decreased effective mass of 
$ 0.85 m_0 $. However, sheets 2 and 3 exhibit increased masses of 
$ 1.87 m_0$ and $ 2.85 m_0$, respectively. These values are in good
agreement with existing Shubnikov-de Haas measurements.\cite{klimm97} 
It has to be noted that the described geometry is not 
the one discussed in conjunction with the ARPES measurements.

\section{Discussion}
\label{discussion}

In general, we found excellent agreement between the experimental
and theoretical Fermi surface contours. Both the ARPES and the
de Haas-van Alphen results confirm the LDA calculations. In particular,
all three Fermi surfaces sheets predicted by theory have been verified
experimentally, with respect to both the shape and the dimensions.

Of course, while from the theoretical side all Fermi surface sheets
can be calculated in one go, the measurements are restricted by the
respective experimental setups and limitations. For instance, only
certain cuts through the Fermi body are usually accessible. In
addition, in the ARPES measurements matrix element effects may also
play a role, which might undermine the experimental detection of
portions of the Fermi surface. This may explain the fact that the
calculated Fermi surface contours of all three sheets can only be
partially observed by ARPES.

While the overall agreement of the results of all three methods is
very good, there still remain small differences on a quantitative
level. These deviations have been included in Tabs.\ \ref{exptab1} and
\ref{exptab2}. To be specific, the deviations of the ARPES Fermi
vectors from the calculated ones amount to several percent but may
go up even to $ 30 $\,\% for some experimental points. Yet, both
negative and positive deviations occur for both the electron- and
hole-like surfaces, precluding systematic errors. This is different
for the de Haas-van Alphen measurements. All Fermi surface contours
resulting from these experiments, including those given in the
literature, are systematically smaller than the calculated ones.
\cite{volskii71,volskii73,volskii75,volskii79,teplinskii80} To
be specific, the electron-like Fermi surface sheet 3 is shifted up
to $ -14 $\,\% in direction of the occupied states, whereas the
hole-like Fermi surface sheets 1 and 2 experience a shift of up to
$ -10 $\,\% in the direction of the unoccupied states. However, we
point out that in general the differences between the experimental
and theoretical data are of a similar size as the differences between
the different experimental techniques.

Nevertheless, all these differences between the experimental and
theoretical results, namely, both the unsystematic deviations of
the ARPES data and the systematically smaller dHvA Fermi surface
contours, still conserve, within the above-mentioned limitations
of the experimental setups, the volume of the Fermi body, which,
by construction, is correctly accounted for by the calculations.
\cite{luttinger60}

For the same reason, non-stoichiometry of the samples can most
likely be ruled out as a source of the observed quantitative
differences. While the phase diagram for Mo-O compounds excludes
a hole-doping of more than $ 1.3 $\,\%,\cite{massalski90} a small
defect concentration of $ 0.12 $\,\% has been detected by Hall effect
measurements.\cite{klimm97} In general, hole-doping causes a
lowering of the Fermi level. According to the band structure shown
in Fig.\ \ref{theofig5}, this is equivalent to a shift of the
Fermi surface contours towards the occupied states. However,
a shift of up to $ -30 $\,\% would necessitate a hole-doping of up
to $ 10 $\,\%. Hence, non-stoichiometry will play only a minor role.

Of course, one might argue, that electronic correlations may affect
the quality of the calculated results. Indeed, our previous
calculations infer the existence of weak electron correlations in
$ {\rm MoO_2} $.\cite{habil,eyert00} According to the present
measurements, effective masses are only slightly enhanced. Yet,
an enhancement may explain the reduction of the band width as it
has been observed in the $ {\bf k}_{\parallel} $ measurements on
the $ {\rm MoO_2} (100)$ surface. However, magneto-transport
measurements performed in this direction do not indicate an
enhanced effective mass.\cite{klimm97,volskii73,volskii75}
Furthermore, as has been demonstrated above part of the experimentally
observed band width reduction could be accounted for by the inclusion
of the full single-particle potential in the present calculations.

Going into more detail, the deviations between the experimental and
theoretical data suggest that, as compared to the experiments, the
calculated Fermi body is slightly shortened parallel to the
$ \bf a $-axis and dilated perpendicular to this axis.
In the directions perpendicular to $ \bf a $ this presumption is
confirmed by the experimental data. In particular, the cuts through
the electron-like ARPES and dHvA Fermi body are smaller than the
calculated results. The assumed shortening parallel to $ \bf a $
is on somewhat weaker grounds. Only the waist of the ``peanut''
could be measured by ARPES, but no experimental points exist at the
top and bottom of the ``peanut''. In contrast, the dHvA measurements
in the direction parallel to $ \bf a $ might be affected by the
underlying ellipsoidal approximation. Yet, our assumption is well
supported by the overall conservation of the volume of the enclosed
Fermi surface as discussed above. Nevertheless, in passing we mention
that these deviation are of the order of a few percent only.

\section{Summary}
\label{summary}

A detailed and comprehensive study of the electronic properties of
$ {\rm MoO_2} $ has been presented. This material has been analyzed
using angle-resolved photoelectron spectroscopy (ARPES), de Haas-van
Alphen (dHvA) measurements, and electronic structure calculations
based on a new full-potential implementation of the augmented
spherical wave (ASW) method.
The Fermi surfaces determined by both kinds of experiments are in very
good agreement with the theoretical predictions. In particular, three
different Fermi surface sheets are correctly identified by all three
approaches. These sheets include an electron-like peanut-shaped Fermi
surface centered around the $ \Gamma $-point, which has a volume of 11\%
of the Brillouin zone, and two smaller hole-like Fermi surfaces both
centered around the Y-point.

As far as they can be detected within the limitations of the
experimental geometries, slight differences between the epxerimental
and theoretical results concern mainly the shape of the three Fermi
bodies. To be specific, from the calculations, the peanut-shaped
Fermi body appears to be slightly shortened and elongated parallel
and perpendicular to the $ {\bf a} $-axis.

With the help of the new calculations, controversies concerning two
additional Fermi surface sheets centered around the Z- and B-point
could eventually be resolved. In particular, the occurence of these
sheets could be attributed to the atomic-sphere approximation, which
was used in our previous work. In contrast, in the new full-potential
ASW calculations, these additional sheets do no longer appear.
Furthermore, the full-potential treatment led to a much improved agreement
between the ARPES results and the calculations as concerns the occupied
band width at the $ \Gamma $-point.

In general, the very good agreement between the results obtained by
the ARPES and dHvA measurements for metallic $ {\rm MoO_2} $ has
important consequences for the investigation of those neighbouring
transition-metal dioxides, which display a metal-insulator transition.
While the Fermi surfaces of the metallic phases of these materials are
not accessible to a low-temperature technique as dHvA experiments, the
good agreement found in the present study implies that ARPES measurements
will be useful to study the Fermi surfaces of the high-temperature
metallic phases.

Furthermore, the good agreement between the different experimental 
techniques and the electronic structure calculations, with the 
deviations between measured and calculated data being of the same 
order as those between the different experimental techniques, support 
our original proposal that these calculations are well suited to study 
the electronic properties of the early transition-metal dioxides. In 
particular, our results are very encouraging concerning the validity 
of our calculations for the $ d^1 $ members $ {\rm VO_2} $ and 
$ {\rm NbO_2} $ and support the interpretation of the metal-insulator 
transitions of the latter two compounds as embedded Peierls-like 
instabilities.

\section{Acknowledgements}
Many useful discussions with K.-H.\ H\"ock, R.\ Horny, G.\ Obermeier,
and S.\ Klimm are gratefully acknowledged. Thanks are due to
S.\ G.\ Ebbinghaus for performing the XRD-measurements.
This work was supported by the Deutsche Forschungsgemeinschaft
through SFB 484 and the European Community through Access to
Research Infrastructures (ARI).

\end{document}